\documentclass[aps,twocolumn,pra,superscriptaddress,longbibliography]{revtex4-2}
\usepackage{graphicx}
\usepackage{dcolumn}
\usepackage{bm}
\usepackage[T1]{fontenc}
\usepackage{textcomp}
\usepackage{amsmath}
\usepackage{mathrsfs}
\usepackage{amssymb}
\usepackage{units}
\usepackage{ulem}
\usepackage{xcolor}
\usepackage[colorlinks,bookmarks=false,citecolor=blue,linkcolor=blue,urlcolor=blue]{hyperref}
\usepackage{soul}
\usepackage{color}

\newcommand{\mrc}[1]{\textcolor{black}{#1}}
\newcommand{\mrb}[1]{\textcolor{black}{#1}}

\begin{document}

\title{\mrc{Emission of time-ordered photon pairs from a coherently-driven Kerr microcavity}}

\author{Ferdinand Claude}
\affiliation{Majulab International Research Laboratory, French National Centre for Scientific Research, National University of Singapore, Nanyang Technological University, Sorbonne Universit\'e, Universit\'e C\^ote d'Azur, 117543 Singapore}
\affiliation{Centre for Quantum technologies, National University of Singapore, 117543 Singapore}

\author{Yueguang Zhou}
\affiliation{Majulab International Research Laboratory, French National Centre for Scientific Research, National University of Singapore, Nanyang Technological University, Sorbonne Universit\'e, Universit\'e C\^ote d'Azur, 117543 Singapore}
\affiliation{Centre for Quantum technologies, National University of Singapore, 117543 Singapore}

\author{Sylvain Ravets}
\affiliation{Universit\'{e} Paris-Saclay, CNRS, Centre de Nanosciences et de Nanotechnologies, 91120 Palaiseau, France}

\author{Jacqueline Bloch}
\affiliation{Universit\'{e} Paris-Saclay, CNRS, Centre de Nanosciences et de Nanotechnologies, 91120 Palaiseau, France}

\author{Martina Morassi}
\affiliation{Universit\'{e} Paris-Saclay, CNRS, Centre de Nanosciences et de Nanotechnologies, 91120 Palaiseau, France}

\author{Aristide Lema\^itre}
\affiliation{Universit\'{e} Paris-Saclay, CNRS, Centre de Nanosciences et de Nanotechnologies, 91120 Palaiseau, France}

\author{Alberto Bramati}
\affiliation{Laboratoire Kastler Brossel, Sorbonne Universit\'e, CNRS, ENS-PSL Research University, Coll\`ege de France, 4 Place Jussieu, 75005 Paris, France}

\author{Anna Minguzzi}
\affiliation{Univ. Grenoble Alpes, CNRS, LPMMC, 38000 Grenoble, France}

\author{Iacopo Carusotto}
\affiliation{Pitaevskii BEC Center, INO-CNR and Dipartimento di Fisica, Universit\`{a} di Trento, via Sommarive 14, I-38123 Trento, Italy}

\author{Ir\'en\'ee Fr\'erot}
\affiliation{Laboratoire Kastler Brossel, Sorbonne Universit\'e, CNRS, ENS-PSL Research University, Coll\`ege de France, 4 Place Jussieu, 75005 Paris, France}

\author{Maxime Richard}
\affiliation{Majulab International Research Laboratory, French National Centre for Scientific Research, National University of Singapore, Nanyang Technological University, Sorbonne Universit\'e, Universit\'e C\^ote d'Azur, 117543 Singapore}
\affiliation{Centre for Quantum technologies, National University of Singapore, 117543 Singapore}


\maketitle

\mrc{\textbf{Weakly-interacting many-body systems possess remarkable quantum properties that are essential components of quantum technologies, and constitute a topic of fundamental interest. Here we show that in a solid-state nonlinear microcavity embedding discrete modes of exciton-dressed photons, we can isolate a single eigenmode of quantum fluctuations from the much brighter coherent fraction of the field. In this regime, we perform frequency- and time-resolved correlations measurements between photons on the red and blue side of the fluctuations spectrum. When the average number of fluctuation quanta is smaller than one, we observe the formation of large pairwise time-ordered correlations: red photon first and blue photon second. We show that this peculiar time-ordering correlation emerges spontaneously from the interplay between frequency-resolved detection, and the non-trivial internal quantum structure of the elementary fluctuations.}}


\begin{figure*}[htpb!]
\includegraphics[width=\textwidth]{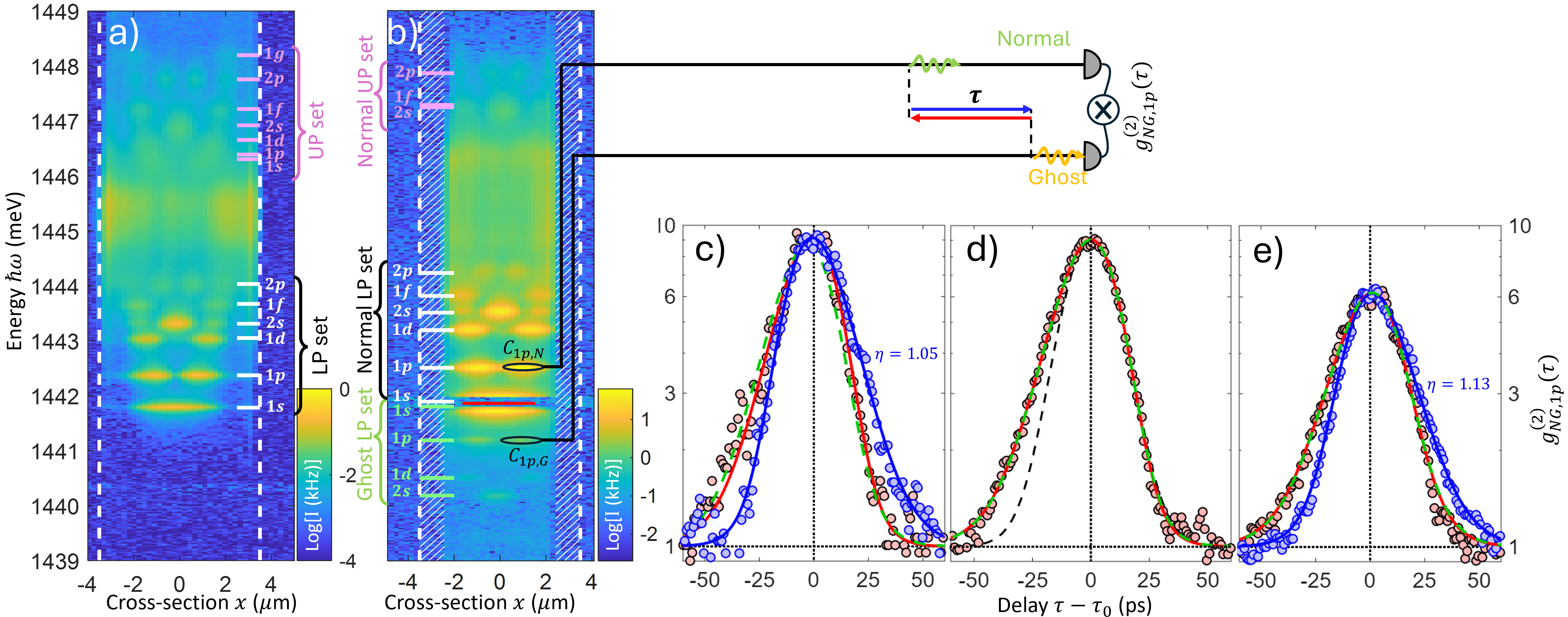}
\caption{\mrc{\textbf{Measurement of time-ordered correlations between normal and ghost photons} Spatially-resolved spectroscopy of a $7\,\mu$m micropillar of (a) photoluminescence $I_{PL}(x,\omega)$ in the low-intensity regime, (b) the Bogoliubov fluctuation photoemission $I_{b}(x,\omega)$ under resonant CW drive in the $1s$ LP mode (b). The photon count rate is plotted in logarithmic color scale. The micropillar edges are shown as vertical dashed lines. The cross hatched region in (b) is rejected by a spatial filter. The calculated micropillar lower (LP) and upper (UP) polariton levels in the linear regime are shown in (a) with their mode labels as white and pink dashes respectively. The calculated energies of the LP Bogoliubov set of modes, Normal and Ghost components, as well as the Normal component of the UP Bogoliubov set of modes, are shown as dashes and labeled accordingly in (b). $C_{N,1p}$ ($C_{G,1p}$) indicate the $(x,\hbar\omega)$ areas from where the photons are collected by the SNSPDs (sketched as the gray detectors).(c-e) Measured (round symbols) normal-ghost photon correlation function $g_{NG,1p}^{(2)}(\tau)$ in a $6\,\mu$m (c), a $7\,\mu$m (d), and an $8\,\mu$m micropillar (in vertical log scale). The solid (dashed) lines are fit with the relaxed (full) model. In the blue curves, the Normal and Ghost detectors have been swapped. Due to statistical variations, the blue curve is rescaled by a factor $\eta-1=13\%$ in (c) and $\eta-1=5\%$ in (e), for easier comparison between swapped and not swapped data (the plotted rescaled correlation is $\eta[g^{(2)}(\tau)-1]+1$). The dotted black curve in (d) shows a fit of the $\tau\geq 0$ side of $g_{NG,1p}^{(2)}(\tau)$, using the symmetric ($\xi=0$) relaxed model. In (c-e), $\tau_0(\omega_{N,G}) \in[-15,15]\,$ps is a small unknown delay due to optical path length dependence on $\omega_{N,G}$.}}
\label{fig:fig1}
\end{figure*}

\mrb{Systems of strongly interacting bosons are characterized by large many-body correlations that give rise to some of the most intriguing quantum states of matter. Remarkably, the weakly interacting regime also possesses a rich and diverse set of quantum properties and phenomena that have proven essential in both the science of quantum fluids and quantum technologies. As we show in this work, this set is far from fully explored, and more quantum resources can be found in unexplored regions of the parameter landscape.}

\mrb{Weakly interacting bosonic systems are best understood as consisting of two distinct contributions: a large coherent field $\psi_0$, and a surrounding cloud of quantum fluctuations. This picture emerges upon linearizing the theory around the steady-state solution, resulting in a quadratic Hamiltonian (involving non-zero anomalous terms) for the description of the fluctuations dynamics \cite{Millburn_book} (see Methods section eq.(\ref{eq:H_f})). In the context of Gaussian states of light, these fluctuations are typically represented in the photon basis, which is better suited to handle observables based on field quadratures. In the context of ultracold atoms, the fluctuations are better known in their diagonal basis, in which they are identified as  \textit{Bogoliubov} fluctuations \cite{Bogoliubov_1947,Pitaevskii_2003}. The two representations are rigorously equivalent \cite{Millburn_book}, but the latter provides a clearer physical picture for observables involving a few fluctuation quanta, like in the present work.}

\mrb{Irrespective of the chosen representation, the quantum dynamics of these fluctuations is at the basis of many paradigmatic microscopic and macroscopic quantum phenomena in various physical systems, such as the emergence of superfluidity in quantum fluids of ultracold atoms~\cite{Steinhauer_2002}, and in microcavities of exciton-polaritons (polaritons)~\cite{Amo_2009,Carusotto_2013}), or the quantum depletion mechanism \cite{Lopes_2017} that fixes a nonzero lower bound to the fluctuations. These fluctuations also govern the non-classical features characterizing the widely known family of Gaussian states in ultracold atoms \cite{Esteve_2008}, and in nonlinear optical systems ~\cite{Andersen_2016}, where they constitute an essential resource for quantum photonic technologies~\cite{Acernese_2023,Yokoyama_2013,Usenko_2026}.}


\mrb{Recently, the focus has moved towards the microscopic quantum features characterizing these fluctuations, with for instance the demonstration in polaritonic systems of a weak quantum blockade regime \cite{Munoz_2019,Delteil_2019,Scarpelli_2024}. Indeed, the Bogoliubov picture shows that a quantum of fluctuation, i.e. an elementary excitation, has a peculiar internal quantum structure, which consists in a quantum superposition of the creation and annihilation of an excited particle, out of the coherent fraction. This structure, sometime more concisely described as a particle and hole superposition \cite{Pitaevskii_2003}, is formally expressed by the Bogoliubov transformation $\hat \beta^\dagger=u\hat a^\dagger - v\hat a$, where $\hat a^\dagger$ and $\hat \beta^\dagger$ are respectively the particle (e.g. photon or atom) and Bogoliubov-excitation creation operators, and $(u,v)$ are the Bogoliubov coefficients. This results for instance in the emergence of a negative-frequency density-of-state in the excitation spectrum (the so-called Ghost component), in addition to the positive-frequency density of states (the Normal Component), that has been observed in ultracold atom systems \cite{Steinhauer_2002, Vogels_2002} and in quantum fluids of light \cite{Utsunomiya_2008,Zajac_2015,Fontaine_2018,Pieczarka_2020,Pieczarka_2021,Claude_2022,Frerot_2023} (See Section \ref{sec:theory} for a formal definition of Bogoliubov's Normal and Ghost components). At the individual quantum level, this structure is also expected to result in pair-correlations \cite{Busch_2014}, as was reported recently in an ultracold atom system \cite{Tenart_2021,Gondret_2025}. Interestingly, in the latter experiment, a time-ordering character was likely present, but it was not anticipated nor observable due to a non-frequency-resolved detection.}

\mrb{In this work, we use a polaritonic microcavity featuring a few well-resolved discrete modes to isolate and investigate a single Bogoliubov fluctuations mode. After rejecting the mean-field, we perform frequency- and time-resolved photon correlation measurements between the ghost and normal components of this fluctuations mode, and we find that whenever the average excitation number of the Bogoliubov field is smaller than one, the correlation becomes large and acquires a strong time-ordered character. This unique feature results from the fact that in this single fluctuations mode configuration, which is different from that used in spontaneous parametric emission of correlated pairs (see SM section III.C), the emission inherits the full particle-hole quantum structure constitutive of Bogoliubov's elementary excitations. }

\section{Experiment}

\mrb{The systems we are investigating are spatially-confined micropillar microcavities of $6\,\mu$m to $8\,\mu$m diameters. These structures feature well split discrete transverse cavity mode resonances, in which the photons are dressed by excitons (bound electron-hole pairs) into polaritons \cite{Carusotto_2013} at cryogenic temperatures $T\in[4,20]\,K$, in order to achieve sizable narrow-band two-body interactions (cf. Methods section A for details on the system). This diameter range optimizes the lower polariton (LP) modes $1s$-$1p$ frequency splitting $\Delta_{1s-1p}$ (see Fig.\ref{fig:fig1}.(a) for the mode labels) for our correlation measurement. Indeed, we need a comparable magnitude of $\Delta_{1s-1p}$ and $g|\psi_0|^2$, where $g$ is the nonlinearity strength and $\psi_0$ the mean-field (see Methods Eq.(\ref{eq:H_0})), to ensure that the $1p$ mode undergoes a significant Bogoliubov transformation (i.e. $|v/u|^2\not\ll 1$), when the $1s$ mode is coherently driven. And we also need $\Delta_{1s-1p} \gg \gamma_{lp,1s}=45\mu$eV$/\hbar$ ($7\mu$m micropillar) in order to minimize the $1s-1p$ spectral overlap ($\hbar\gamma_j$ is $j$ mode's full-width at half-maximum, i.e. FWHM).}

\mrb{To study these micropillar resonators, we developed an experimental setup providing spatially-resolved ultra-fast frequency and time two-photon correlation measurements and spectroscopic measurements capabilities (see detailed description in SI section I.A). We first characterize the system's linear property by performing a spatially-resolved photoluminescence spectroscopy $I_{pl}(\omega,x)$ measurement under low-intensity non-resonant CW excitation. The result is shown in Fig.\ref{fig:fig1}.(a) for a $7\,\mu$m diameter micropillar with the emission intensity in color log-scale. We observe two sets of discrete modes: the upper (UP) and lower (LP) polariton sets, as expected in polaritonic microcavities. The lowest energy (LP) transverse modes meet the condition $\Delta_{1s-1p} \gg \gamma_{lp,1s}$. We find that both the measured modes frequencies $\omega_j$, and spatial profile ($x$-coordinate), agree quantitatively with a model of cylindrically confined microcavity in the strong coupling regime (see white and pink dashes with corresponding mode labels in Fig.\ref{fig:fig1}.(a), see Methods section B.1 for the model).}

\mrb{We then drive quasi-resonantly the LP $1s$ mode by setting the laser frequency to $\hbar\delta_{las}=\hbar\omega_{las}-\hbar\omega_{lp,1s}=+90\,\mu$eV. In this blue-detuned regime, owing to the positive nonlinearity $g$, the intracavity average photon number $|\psi_0|^2$ is bistable with respect to the driving laser intensity $I_{las}$ \cite{Carusotto_2013}. We work in the high photon-number branch of the hysteresis curve, where the condition $\Delta_{1s-1p}\simeq g|\psi|^2$ is achieved, and far above the bistability window (that lies typically within $150<I_{las}<190\,\mu$W ), at $I_{las}\simeq 500\,\mu$W, to ensure long term stability (see SI section I.C for a typical hysteresis curve and working point).}

\mrb{The resulting photoemission passes through a custom-made high-rejection narrow-band filter to remove entirely the mean-field component, and keep only the fluctuations at all frequencies $|\hbar\omega|\geq\hbar\omega_{las}+75\,\mu$eV \cite{Frerot_2023} (see SI section I.A). Fig.\ref{fig:fig1}.(b) shows such a spatially-resolved fluctuations-only emission spectrum $I_b(\omega,x)$. It shows a set of discrete fluctuations modes above the laser frequency (the Bogoliubov LP and UP modes' Normal components), of frequencies and spatial profile close to the modes in the linear regime, yet shifted by the Bogoliubov transformation (white and pink dashes in Fig.\ref{fig:fig1}.(b)). It shows also a new set of modes, mirror-symmetric with respect to the laser frequency, that correspond to the LP Bogoliubov modes' Ghost components (green dashes in Fig.\ref{fig:fig1}.(b)). The Normal and Ghost components of the $1s$, $1p$, $1d$ and $2s$ LP-Bogoliubov modes are spectrally well-split from each other: we can thus isolate the Normal-Ghost frequency pair of a single one (e.g. the $1p$ mode) for correlation measurements. To check that this fluctuation spectrum is consistent with the Bogoliubov picture, we developed, and numerically solved a driven-dissipative Bogoliubov model for cylindrical resonators (see Methods section B.2). The resulting fluctuations resonances and mode labels are shown as the colored dashes and labels in Fig.\ref{fig:fig1}.(b), and found to be in quantitative agreement with the measurement.}

\mrb{We now proceed to measure the time-resolved Normal-Ghost pair correlations of the $1p$ fluctuation mode, that best fits the $1s-1p$ modes splitting condition discussed earlier. The input ends of the superconducting single photon detectors (SNSPDs) fibers are optically conjugated with the spectrometer output focal plane, in order for each of them to select a small frequency-position collection areas $C_{1p,N}$ and $C_{1p,G}$ (indicated by the two black ellipses in Fig.\ref{fig:fig1}.(b). $C_{1p,N}$ ($C_{1p,G}$) is thus positioned to collect the Normal (Ghost) frequency of the $1p$ Bogoliubov mode, and spatially, on a bright antinode of the mode pattern (see SI Section I.B for details on the photons collection). The collection spectral bandwidth (the short axis of $C_{1p,NG}$ in Fig.\ref{fig:fig1}.(b)) is tuned to $\Gamma_{bw}\simeq 120\,\mu$eV, i.e. twice larger than $\hbar\gamma_{1p}\simeq 55\,\mu$eV (FWHM), providing a good frequency/time-resolution tradeoff. The collected spatial area (the long axis of $C_{1p,NG}$ in Fig.\ref{fig:fig1}.(b)) corresponds to $\sim 20\%$ of the total mode area. The Normal-ghost correlations $g^{(2)}_{NG,1p}(\tau)$ is obtained by counting photon pair events involving a delay $\tau\pm\Delta\tau/2$ (where $\Delta\tau=1\,$ps is the delay time-bin size; see a raw measured events histogram in SI section I.A Fig.S1), and by normalizing the result using the SNSPD's count rates (see SI section I.B). Accounting for the jitter sources, the delay is determined with an accuracy ($1\sigma-$confidence interval) of $7.3\,$ps.}

\mrb{Measured correlation functions $g^{(2)}_{NG,1p}(\tau)$ are shown in Fig.\ref{fig:fig1}.(c-e) at the lowest cryostat temperature $T=4.0\,$K for the $6\,\mu$m (Fig.\ref{fig:fig1}.(c)), $7\,\mu$m (Fig.\ref{fig:fig1}.(d)), and $8\,\mu$m (Fig.\ref{fig:fig1}.(e)) micropillar. In these plots $\tau>0$ ($\tau<0$) corresponds to events with the normal (ghost) photon detected first. These data show strong pair correlations within a delay interval comparable to $\gamma_{1p}^{-1}$, and of magnitude ${\rm max}[g^{(2)}_{NG,1p}(\tau)]=9.7 \pm 0.8$, ${\rm max}[g^{(2)}_{NG,1p}(\tau)]=9.0 \pm 0.3$, and ${\rm max}[g^{(2)}_{NG,1p}(\tau)]=6.2 \pm 0.3$, respectively (reminding that uncorrelated pairs yield $g^{(2)}(\tau))=1.0$). These correlation peaks also all exhibit a pronounced $\pm\tau$ asymmetric shape. For visual comparison, a symmetric lineshape matching the right side of the measured $g^{(2)}_{NG,1p}(\tau)$ is shown in Fig.\ref{fig:fig1}.(d), and we repeated the measurement for the $6\,\mu$m and $8\,\mu$m micropillar with the detectors' Normal-Ghost assignment swapped (blue symbols).}

\mrb{As this asymmetric correlation is central to this work, we performed several verification measurements, to make sure that it does not stem from an experimental artifact (see a detailed summary in SI section II.A.3). This verification includes for instance the measurement, in the same condition, of the $1p$-Normal/$1p$-Normal autocorrelation function $g^{(2)}_{NN,1p}(\tau)$. The obtained correlation peak is found to be fully symmetric, with a magnitude ${\rm max}[g^{(2)}_{NN,1p}(\tau)]=1.98 \pm 0.13$, which is in quantitative agreement in both shape and magnitude with theoretical expectations (see SI section II.A.1 for a detailed analysis). The asymmetry that we observe is thus a characteristic of the emission quantum statistics: the magnitude of the pair correlations depends on the order in which the Normal and Ghost photons constituting the pair, are emitted. We explain thereafter, both physically and quantitatively, how this time-ordering emerges out of a single mode of Bogoliubov fluctuations.}



\section{Physical principle and Theory}
\label{sec:theory}

\begin{figure}[tp]
\includegraphics[width=0.9\columnwidth]{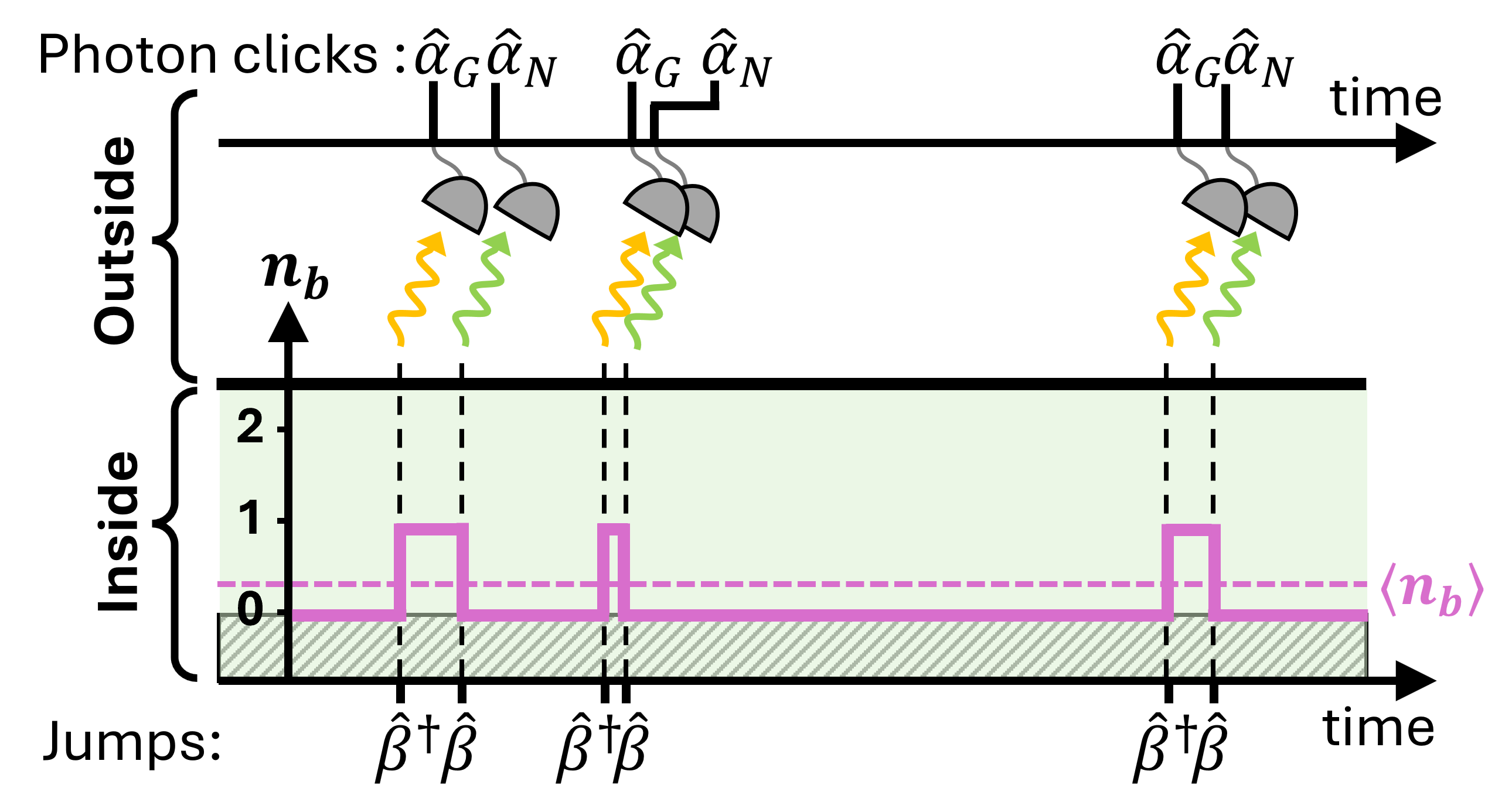}
\caption{\textbf{Qualitative time-evolution of the number of excitation quanta $n_b(t)$ within a Bogoliubov field, in the regime $\langle \hat n_b\rangle=\langle \hat{\beta}^\dagger\hat{\beta}\rangle\ll 1$}. The Bogoliubov field (purple thick line) has a high probability to be in its ground state $n_b=0$ at any time, so that the subtraction of one quantum from the field, accompanied by its green photon witness, is necessarily preceded by an excitation of the field, accompanied by its yellow photon witness, while the opposite order yields no correlation. "inside" ("outside") refers to the intra- (extra-) cavity medium. The Bogoliubov field only exists "inside". $\hat \alpha_j$ describe the green ($j=N$) and yellow ($j=G$) photon operators acting on the detectors.}
\label{fig:fig1p5}
\end{figure}

In the experimental conditions of this work, a polaritonic microcavity can be formally considered as a Kerr-nonlinear photonic microcavity \cite{Carusotto_2013}. We thus describe our system as such, and consider that it features a single Bogoliubov mode of frequency $\omega_B$, and operator $\hat \beta$ (see Methods section C.1). The photonic fluctuation equation of motion can be simply derived and Fourier transformed, leading to (see Methods section C.1 Eq.(\ref{eq:EOM_a_of_beta_andt})) 
\begin{equation}
    \tilde{a}(\omega) =  u \hat{\beta} \delta(\omega-\omega_B)+ 
    v \hat{\beta}^\dagger \delta(\omega+\omega_B),
    \label{eq:NG_2frequencies}
\end{equation}
where $\hat a(\omega_B)=\int_{\omega_B\pm\Gamma_{bw}/2} \mathrm{d}\omega'\tilde{a}(\omega')$, $\Gamma_{bw}$ is the detection bandwidth, and where, owing to the linear input-output relation (see Methods section C.1), $]\hat a(+\omega_B)$ ($\hat a(-\omega_B)$) is directly proportional to the photon operators acting on the Normal-photon- (Ghost-photon)-coupled detector. One then obtains two remarkable identities,
\begin{eqnarray}
      \hat a(\omega_B)   &=&  u \hat{\beta} \equiv \hat{a}_N \\       
      \hat a(-\omega_B)  &= & v \hat{\beta}^\dagger \equiv \hat{a}_G.
    \label{eq:a_of_b}
\end{eqnarray}
that establishes that, as illustrated in Fig.\ref{fig:fig1p5}, the detection of a Normal (Ghost) photon $\hat a_N$ ($\hat a_G$) is an unambiguous witness of the fact that a quantum of fluctuation has been subtracted from (added to) the Bogoliubov field. This is the key mechanism behind the emergence of time-ordering in our measurement, as will be established further. 

The pair detection rate involved in our correlations measurements read $\mathcal{I}^{(2)}_{NG}=\langle \hat{a}^\dagger_N \hat{a}^\dagger_G \hat{a}_G \hat{a}_N\rangle$  or $\mathcal{I}^{(2)}_{GN}=\langle \hat{a}^\dagger_G \hat{a}^\dagger_N \hat{a}_N \hat{a}_G\rangle$, depending on whether the Normal or Ghost photon is detected first. Using the identities derived above, these rates can be written in the Bogoliubov basis as $\mathcal{I}^{(2)}_{NG}\propto \langle \hat{\beta}^\dagger\hat{\beta}\hat{\beta}^\dagger\hat{\beta} \rangle$ and $\mathcal{I}^{(2)}_{GN}\propto \langle \hat{\beta}\hat{\beta}^\dagger\hat{\beta}\hat{\beta}^\dagger \rangle$, which hence have a different magnitude. Physically, owing to the Bogoliubov operators ordering in each expression, detecting a photon pair with a leading Normal photon (NG) or a leading Ghost photon (GN) thus identifies the occurrence of two different events within the Bogoliubov quantum field: as illustrated in Fig.\ref{fig:fig1p5}, the detection of a NG (GN) pair identifies the occurrence of a stochastic transient suppression (addition) of one elementary excitation in the field. 

Completing this derivation with the help of Wick's theorem yields
\begin{eqnarray}
  \mathcal{G}^{(2)}_{NG} \simeq 1+\frac{n_{b}}{1+n_{b}}  \label{eq:ideal_g2_NG}\\
  \mathcal{G}^{(2)}_{GN}  \simeq 1+\frac{1+n_{b}}{n_{b}} \label{eq:ideal_g2_GN},
\end{eqnarray}
where $\mathcal{G}^{(2)}_{q}$ is the normalized second-order Normal-Ghost photon pair correlation, $n_b=\langle\hat{\beta}^\dagger\hat{\beta}\rangle$ is the average number of Bogoliubov excitations, and where we have assumed that the Bogoliubov field is in a Gaussian state and that $\langle \hat{\beta}\hat{\beta}\rangle\propto\gamma/\omega_B\simeq 0$ (see SM section II).

\mrb{There are two interesting regimes in these expressions. When $n_b\gg 1$, the Bogoliubov field is high in the excitation ladder, so that the transient suppression or addition of an excitation quantum is equally likely. The correlations thus do not depend on the photons time-ordering and converge to $\mathcal{G}^{(2)}_{GN}=\mathcal{G}^{(2)}_{NG}=2$. The regime $n_b<1$, which is illustrated in Fig.\ref{fig:fig1p5}, is the most interesting: owing to the rarity of the fluctuations, and to the high probability of the system to be found in the fluctuations ground state, the transient suppression of an excitation (NG time ordering) is, most of the time, impossible, while the transient creation (GN time ordering) is a strongly dominating event. This is directly reflected by the two correlations as, when $n_b$ goes to zero, $\mathcal{G}^{(2)}_{NG}\simeq 1$, while $\mathcal{G}^{(2)}_{GN} \rightarrow \infty$. From the fluctuation photoemission point-of-view, this regimes means that the emission of a Normal photon is always preceded by the emission of a Ghost photon, while the emission of a Ghost photon carries no information on the prior emission of a normal photon. This time-ordering mechanism is a unique characteristics of the single Bogoliubov mode, and appears only when $n_b<1$.}

\mrb{Note that these result can be all reformulated in the photon fluctuations basis using $n_a=\langle \hat a^\dagger \hat a\rangle$, and the two-photon coherence $c_a=\langle \hat a \hat a\rangle$ that fully characterizes any gaussian states. In this picture, the usual fluctuations squeezing parameters $(r,\phi)$ are related to the Bogoliubov coefficients according to $u=\cosh(r)$, $v=e^{-2i\phi}\sinh(r)$ \cite{Millburn_book}.}

With the physical picture clarified, we developed this model further to achieve a quantitative comparison with the measurements. We did so by formulating it into an input-output Bogoliubov theory, which is able to handle finite values of the two-photon delay $\tau$, and of the detectors bandwidth $\Gamma_{bw}$. In the limit where $\langle\hat{\beta}\hat{\beta}\rangle\simeq 0$, and $\omega_B \gg (\Gamma_{bw},\gamma)$, we find that the delay-dependent Normal-Ghost correlation function $g^{(2)}_{NG}(\tau)$ can be explicitly derived as (see Methods section C.2)

\begin{eqnarray}
    && g^{(2)}_{NG}(\tau) = 1 + A\frac{\Gamma_{bw}^2}{(\Gamma_{bw}-\gamma_b)^2}\times \nonumber \\ 
    && \left[\big(1-s_\tau\xi \big)e^{-\gamma_b|\tau|}-\bigg(\frac{\gamma_b}{\Gamma_{bw}}-s_\tau\xi\bigg)e^{-\Gamma_{bw}|\tau|}\right]^2,
    \label{eq:g2_of_tau_single_mode}
\end{eqnarray}
where 
\begin{eqnarray}
    & A &=1+[4n_b(n_b+1)]^{-1} \label{eq:A_of_n}\\ 
    & \xi&=(2n_b+1)^{-1}.
    \label{eq:xi_of_n}
\end{eqnarray}
In this expression, $s_\tau$ is the sign of $\tau$, which is positive (negative) when the normal (ghost) photon is detected first. $A$ is a characteristic correlation amplitude, and $\xi$ is the magnitude of the time-ordering-induced asymmetry. $\xi=0$ means a symmetric $g^{(2)}_{NG}(\tau)$, while $\xi=1$ ($\xi=-1$) means a maximally asymmetric $g^{(2)}_{NG}(\tau)$, with ghost-photon-first (Normal-photon-first) order. Consistently, in the regime where $\omega_B \gg \Gamma_{bw} \gg \gamma_b $, and for Normal-Ghost photons delays $\Gamma_{bw}^{-1}\ll\tau\ll \gamma_b^{-1}$, $g^{(2)}_{NG}(\tau)$ matches $\mathcal{G}^{(2)}_{NG}$ for $\tau>0$ and $\mathcal{G}^{(2)}_{GN}$ for $\tau<0$. Note also that Eq.(\ref{eq:g2_of_tau_single_mode}) does not depend on the mode spatial pattern (in this single mode case), the latter being canceled out by the normalization factor.

\mrc{Remarkably, this expression depends only on $n_b$, and not on the microscopic processes populating $n_b$, namely: the coupling to the extra-cavity vacuum \cite{Busch_2014}, the thermal solid-state phonons that couple to polariton via their excitonic fraction \cite{Frerot_2023}, or the electronic noise \cite{Stepanov_2019}. Like in the simpler model Eqs.(\ref{eq:ideal_g2_NG}-\ref{eq:ideal_g2_GN}), the key feature of this expression is that $g^{(2)}_{NG}(\tau)$ becomes increasingly asymmetric, i.e. time-ordered, for decreasing $n_b\leq 1$, and correspondingly its peak magnitude increases, approximately as $(n_b+1)/n_b$. Within this model, these two observables are locked to each other by a universal formula 
\begin{equation}
A=(1-\xi^2)^{-1},
\label{A_of_alpha}
\end{equation}}
which is thus an additional characteristic feature of this phenomenon. 

\section{Analysis}

\mrb{By fitting our measurements with Eq.(\ref{eq:g2_of_tau_single_mode}) and not enforcing Eq.(\ref{A_of_alpha}) (the 'relaxed' model), we can extract the characteristic amplitude $A$, and asymmetry $\xi$ from the measured correlation function, while using Eq.(\ref{eq:g2_of_tau_single_mode}) and enforcing Eq.\eqref{A_of_alpha} (the 'full' model) tests the agreement with the theory. The results using the relaxed model is shown for the three $g^{(2)}_{NG,1p}(\tau)$ measurement of Fig.\ref{fig:fig1}.(c-e) as the solid line, yielding $\xi_M=1.0 (-0.08,+0.0)$, $\xi_M=1.0 (-0.03,+0.0)$, and $\xi_M=0.93 (-0.06,+0.07)$ respectively. These asymmetry values $\xi$ have the expected positive sign (negative sign when the detectors are swapped), corresponding to a leading ghost photon, and they match the maximal theoretical asymmetry (time-ordering) magnitude $\xi=1$. Fits to the data with the full theory are shown on the same plots as the dashed green line. A strong agreement is found, providing an estimate of the number of Bogoliubov excitations quanta in the $1p$ mode of $n_b=0.046\pm 0.002$, $n_b=0.044\pm 0.001$, and $n_b=0.068 \pm 0.002$, showing -- consistently with the observed strong asymmetry -- that at this low temperature ($T=4.0\,$K), the $1p$ Bogoliubov mode is in the $n_b\ll 1$ time-ordered correlation regime. }

\begin{figure}[t]
\includegraphics[width=\columnwidth]{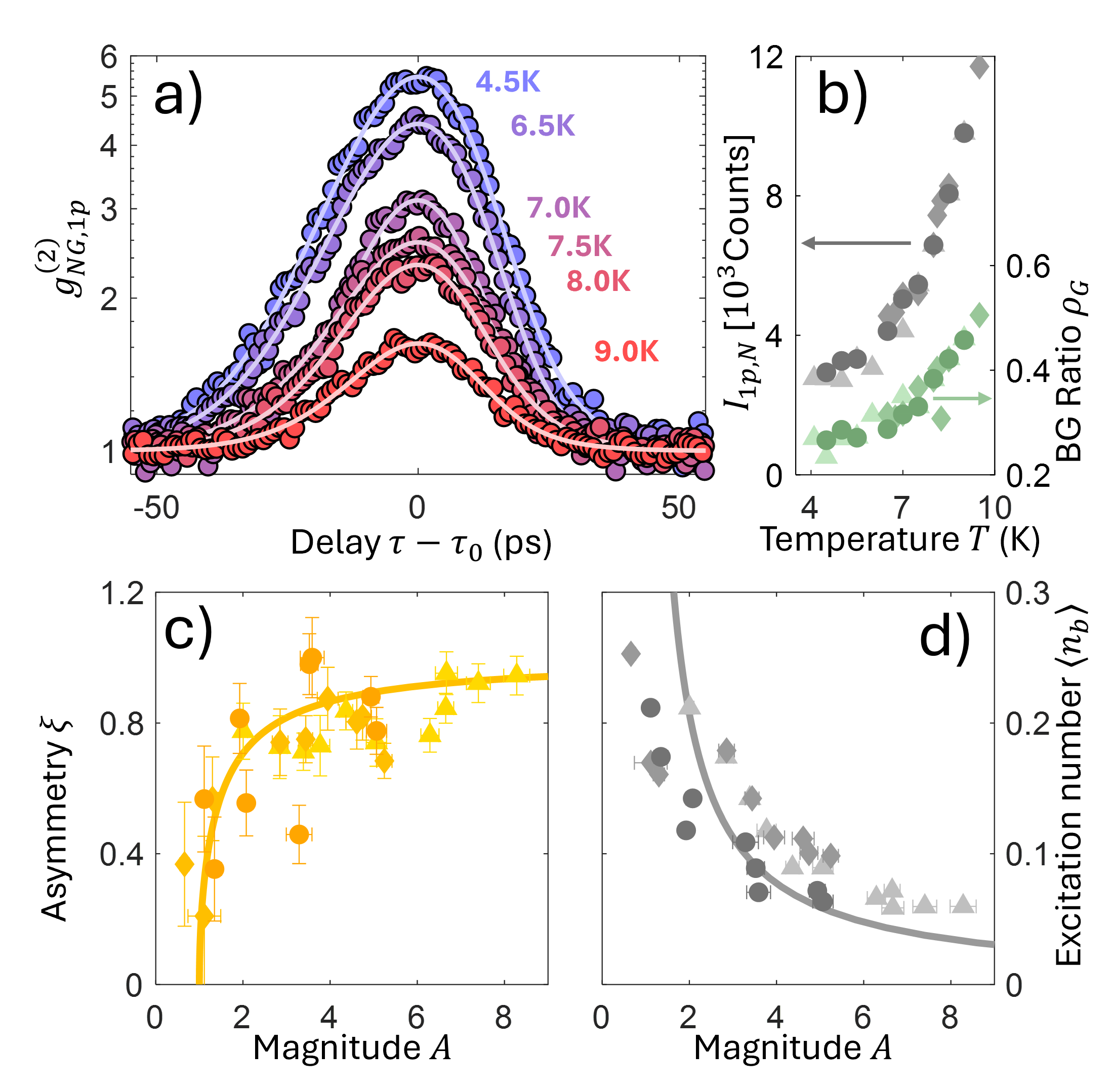}
\caption{\textbf{Correlation amplitude $A$ and asymmetry $\xi$ of the $1p$ Bogoliubov mode emission as a function of temperature T --} (a) Measurement of $g_{1p}^{(2)}(\tau-\tau_0)$ as a function of the micropillar temperature (blue to red symbols for temperatures increasing from $4.4\,$K to $9\,$K). $\tau_0(\omega_{N,G}) \in[-15,15]\,$ps is a small unknown delay offset due to optical path length dependence on $\omega_{N,G}(T)$. The solid lines are fits to the data using the relaxed lineshape model. b) Measured emission intensity $I_{1p,N}$ of the normal $1p$ Bogoliubov mode (left axis), and proportion $\rho_G$ of background emission at the $1p$ Ghost frequency (right axis), as a function of temperature. (c) Measured (symbols) correlation asymmetry $\xi_M$ as a function of the measured correlation characteristic amplitude $A_{Mc}$, corrected from the uncorrelated background photons. The theoretical $\xi(A)$ is plotted as a solid line. (d) Measured (symbols) $1p$ number $n_{b,M}$ of Bogoliubov excitation as derived from (b) as a function of $A_{Mc}$. The theoretical $n_b(A)$ is plotted as a solid line. In (b,c,d), the three color shades and symbols (circles, diamonds, triangles) corresponds to three separate experimental runs, the data in (a) come from the first run.}
\label{fig:fig2}
\end{figure}



\mrc{According to Eqs.(\ref{eq:A_of_n},\ref{eq:xi_of_n}), increasing $n_b$ should decrease both $\xi$ and $A$, and both quantities should remain linked with each other by Eq.\eqref{A_of_alpha}. We then take advantage of the fact that in polaritonic systems, the Bogoliubov excitations are coupled to the thermal bath of solid-state photons \cite{Frerot_2023}, and $n_b$ can thus be increased by increasing the microcavity temperature. We thus perform sets of temperature-dependent measurements of $g^{(2)}_{NG,1p}(\tau)$ between $T=4.0\,$K to $20\,$K. The result is shown for the $7\,\mu$m micropillar in Fig.\ref{fig:fig2}.(a). The data exhibit the expected trend: for increasing temperature, the peak correlation value $\max[g^{(2)}_{NG,1p}(\tau)]$, and hence $A$, decreases, while the asymmetry gets smaller.}

\mrc{To test if this trend follows the characteristic Eq.(\ref{eq:xi_of_n}), we fit the data using the relaxed model and hence obtain a set of $(\xi,A)[T]$ pairs. For this analysis to be quantitatively meaningful, the fitted correlation amplitude is additionally corrected to account for a weak background of photons that contributes at the $1p$ Ghost frequencies, but come from the $1s$ spectral tail, and from an unstructured excitonic background. This is done using a separate measurement of the relative background contribution $\rho_G(T)$ shown in Fig.\ref{fig:fig2}.(b) (see SI section II.C for the correction procedure). Since $\rho_G(T)$ depends also on the spatial collection point within the mode, we perform the same temperature-dependent measurement 3 times to estimate the uncertainty coming from the positioning accuracy at the mode anti-node. The data pairs $(A,\xi)[T]$ obtained in this way are plotted in Fig.\ref{fig:fig2}.(c), alongside the theoretical prediction Eq.(\ref{A_of_alpha}) for $A(\xi)$. A solid agreement is found between the two, confirming that the observed correlation parameters $A$ and $\xi$ are governed by the time-ordering mechanism discussed above, over the explored range of $n_b(T)$. }

\mrc{For additional consistency check, we measured the $1p$ Bogoliubov excitation number $n_{b}(T)$ independently from the $\xi$ and Eq.(\ref{A_of_alpha}), using the fact that $n_b(T)= KI_{1p,N}(T)$, where $I_{1p,N}(T)$ is the Bogoliubov emission intensity at the Normal $1p$ frequency, integrated over the narrow emission peak, and that within this temperature range, $K$ is a single unknown constant. The values of $n_b(T)$ obtained in the $7\,\mu$m microcavity $1p$ mode is plotted in Fig.\ref{fig:fig2}.(d), as a function of the correlation amplitude $A(T)$ determined earlier. The pairs $(n_b,A)[T]$ are expected to agree with Eq.(\ref{eq:A_of_n}) to within a single unknown factor $K$ which is left as a free fit parameters. We obtain a reasonable agreement, which allows us to determine that $n_b$ increases from $n_b\simeq 0.05$ at $T=4\,$K to $n_b\simeq 0.25$ at $T=10\,$K. In all the considered range of temperatures, the system thus remains deep in the quantum regime.}

\section{Discussion}

\mrc{In order to put our observations in a broad context, let us highlight other physical systems that exhibit time-ordered photon correlations. Unlike in the present work, most of these systems operate in the strongly nonlinear regime, as they usually involve a quantum emitter. In two-level atoms for instance, coherently driven with a detuned laser, time-ordered photons are observed in the cross-correlation between the opposite sidebands of a Mollow triplet \cite{Aspect_1980}. The time-ordering is due in this case to the fact that blue and red transition within the Jaynes-Cumming ladder do not have the same radiative rates, as the detuned laser results in dressed-state with uneven dipole-photons fraction~\cite{Nienhuis_1993,delValle_2012,Ulhaq_2012,Ng_2022}. This mechanism bears no analogy with the present work. Inverted three-level systems, like the biexciton-exciton cascade in solid-state quantum dots \cite{Moreau_2001} can also produce time-order correlated pairs, that come solely from the underlying fermionicity of the emitter. Note that the notion of cascading can sometime be used to describe zero-delay correlation between photons of different frequencies in a different context \cite{Scarpelli_2024}. Such a correlation however, does not indicate any time-ordering within the detected pair.}

\mrb{Finally, let us mention that the Normal-Ghost photons collected out of a single Bogoliubov mode, are in a fundamentally different state from the photons collected out of a pair of signal-idler modes in a configuration of spontaneous parametric emission, or down-conversion. Due to the resonant condition for parametric scattering, such a configuration is optimal -- and widely used -- for the bright generation of correlated photon pairs \cite{Yuchen_2021}, but due to the signal-idler modes symmetry, these pairs involve no preferred time-ordering. The single-Bogoliubov mode configuration on the other hand, is far from this resonance condition and hence provides a lower pair emission rate; but in return, the pairs acquire this unique time-ordered character (see detailed discussion in SM section III.C).}

\section*{Conclusion}


\mrc{In this work, guided by the Bogoliubov picture on one hand, and owing to the flexibility of polaritonic microcavities on the other hand, we have shown that a single Bogoliubov fluctuations mode, driven in the limit of low excitation quanta number $n_b<1$, and monitored with frequency-resolved observables, generates time-ordered photon pairs: red first, and blue second. We show that this property emerges naturally in the Bogoliubov framework, from the fact that when the fluctuation field is in its ground state, the subtraction of a fluctuation quantum is necessarily preceded by a creation, while the opposite order entails no analogue requirements. A notable outcome of our work is that a large variety of gaussian-states-producing systems, driven into the appropriate regime, are expected to have this capability to generate time-order photon pairs, which constitutes for instance a key resource to prepare time-frequency entanglement \cite{Xavier_2025}.}

\section*{Acknowledgements}
MR, AB, FC and YZ acknowledge financial support from the Centre for Quantum technologies 'Exploratory Initiative program', and from the National Research Foundation via CNRS@CREATE internal grant 'NGAP'. IC acknowledges financial support from Provincia Autonoma di Trento (PAT); from the Q@TN Initiative; from the National Quantum Science and Technology Institute through the PNRR MUR project under Grant PE0000023-NQSTI, co-funded by the European Union -- NextGeneration EU.  SR, AB, AL and MM acknowledge support from the European Research Council under the European Union's Horizon 2020 research and innovation programme through the Grant ARQADIA (grant agreement no. 949730) and under the Horizon Europe programme (project ANAPOLIS, grant agreement no. 101054448), by the RENATECH network and the General Council of Essonne, by the Paris Ile-de-France R\'egion via DIM SIRTEQ and DIM QUANTIP, by the Plan France 2030 through the project QUTISYM ANR23-PETQ-0002, and by the ANR project Ngauge (ANR-24-CE92-0011). AB is member of the Institut Universitaire de France. MR wishes to thank Thomas Volz for stimulating discussions.

\section*{Methods}

\subsection{Micropillar-microcavities structure}
\label{metd:sample}
\mrc{The investigated microcavities are monolithic solid-state semiconductor hetero-structures made of Ga(In,Al)As. Each micropillar features transverse modes of exciton-polaritons, that result from the strong-coupling regime between discrete laterally-confined cavity modes, and the excitonic transition of a single InGaAs quantum well placed at the longitudinal cavity mode antinode \cite{Bloch_1997,Gutbrod_1998,Eldaif_2006}. The polaritonic character is essential to provide an effective two-body interaction, which is large enough to contribute significantly within the intracavity lifetime, and low enough to place the system in the weakly interacting regime in which the Bogoliubov picture is valid. The lower-polariton subset which is used in this work behaves effectively, as discrete Kerr-cavity modes \cite{Carusotto_2013}. The polaritonic regime is achieved upon cooling the system in a cryostat at temperatures in the $T=[4,20]\,$K range. The micropillars diameters range between 6$\,\mu$m and 8$\,\mu$m, with the lowest mode ($1s$) energy in a range between $1441\,$meV to $1445\,$meV, and an effective interaction strength (i.e. Kerr nonlinearity parameter) $U/\gamma_{1s}\simeq 2\%$, and $\hbar\gamma_{1s}\simeq 45\,\mu$eV to $55\,\mu$eV (FWHM).}

\subsection{Multimode theory of nonlinear a 0D micropillar microcavity}

\subsubsection{Polaritonic levels in the linear regime}
\label{metd:lin_levels}

Following an approach similar to that in \cite{Idrissi_2006}, we can describe the conservative part of the system with the exciton-polariton Hamiltonian
\begin{equation}
\begin{split}
\hat{H}_0 = \hbar\sum_j \left[ \omega_{c,j} \hat{a}_j^\dagger \hat{a}_j + \omega_x \hat{b}_j^\dagger \hat{b}_j 
+ \frac{\Omega}{2} \left( \hat{a}_j^\dagger \hat{b}_j + \hat{b}_j^\dagger \hat{a}_j \right) \right] + \hat H_V\\
\end{split}
\end{equation}
where $\omega_x$ is the the excitonic transition frequency, $\Omega$ is the Rabi frequency coupling cavity photons and excitons, and $\omega_{c,j}$ is the cavity resonance of transverse mode $j$. The transverse modes $\hat a_j$ in cylindrical micropillars of radius $R$ ($R=3.5\,\mu$m for this micropillar) are described by the set of eigen-functions $\phi_{n,m}(r,\theta)=\sqrt{A_{n,m}}e^{im\theta}J_m{(z_{n,m}r/R)}$, where $\mathbf{r}=(r,\theta)$ are the spatial coordinate, $j=(n,m)$ are the radial and azimuthal numbers, $A_{n,m}$ is the normalization factor, $J_m(x)$ are Bessel the functions of order $m$, and $z_{n,m}$ is the $n^{th}$ zero of $J_m(x)$. The transverse mode $j$ thus reads $\hat a_j = \int_{r\le R} d^2{\bf r}~ \phi_j({\bf r})\hat a({\bf r})$ with $\hat a({\bf r})$ the cavity-photon field at in-plane position ${\bf r}=(r,\theta)$. We define similarly $\hat b({\bf r})$ the quantum well exciton field, that we also express in the Bessel function basis via the set of operators $\{\hat b_j\}$.

The Hamiltonian term $\hat H_V = \int_{r\le R} d^2{\bf r} ~ h_V(r) \hat b^\dagger({\bf r}) \hat b({\bf r})$, with $h_V(r) = -\hbar a_2(r/R)^2$ describes the excitonic potential within the micropillar, that we determined in a separate micro-spectroscopy analysis. This radial potential accounts for the crystalline strain relaxation at the micropillar edges. We can also express $\hat H_V$ in the Bessel function basis. More specifically, we use the $N=94$ first eigenstates of the bare photonic micropillar microcavity of radius $R$ and refractive index $n_0$, as the representation basis.
 
We thus obtain $\hat H_0$ in a ($188\times 188$) matrix form. Diagonalizing $\hat H_0$ provides the confined polariton levels and eigenstates (cf. modes levels and labels in Fig.\ref{fig:fig1}.(a)), that we can compare (both in terms of energy and cross-section amplitude), with the spatially-resolved measured spectra, like that in Fig.\ref{fig:fig1}.(a). We can thus determine the parameters describing the system, namely: $\{\hbar\omega_x,\hbar\omega_{c0},\hbar\Omega,\hbar a_2\}=\{1445.83 \pm 0.02,1442.078\pm 0.02,3.47\pm 0.02, 1.31 \pm0.03\}\,$meV, where $\omega_{c0}$ is the unconfined cavity mode frequency and $n_0=3.51\pm 0.01$. 

 
\subsubsection{Polaritonic Bogoliubov levels}
\label{metd:bog_levels}
We add a coherent drive $\hat H_D=(F^*e^{i\omega_{\rm las}t} \hat a_{j_0} + {\rm h.c.})$ to $\hat H_0$, where $j_0$ labels the driven mode (for the measurement in Fig.\ref{fig:fig1}.(b), $\hbar\omega_{\rm las}=1441.85\,$meV). We assume that the pump-induced nonlinearities have a negligible effect on the spatial structure of the driven mode. We then add the typical interaction terms known to be relevant to exciton-polaritons \cite{Richard_2025}: $\hat V_{xx} = (\hbar g_x/2)\int_{r\le R} d^2 {\bf r}~ \hat b^\dagger({\bf r}) \hat b^\dagger({\bf r}) \hat b({\bf r}) \hat b({\bf r})$ that describes exciton-mediated scattering, and $\hat V_{\rm sat} = (-\hbar g_s/2) \int_{r\le R} d^2 {\bf r}~ \left[\hat a^\dagger({\bf r}) \hat b^\dagger({\bf r}) \hat b({\bf r}) \hat b({\bf r}) + {\rm h.c.}\right]$ that describe an oscillator saturation mechanism. Our resonantly driven nonlinear micropillar microcavity is described by $\hat H_0 + \hat H_D + \hat V_{xx} + \hat V_{\rm sat}$

To describe the Bogoliubov levels, we first derive the resonantly driven mode steady-state solution for the excitonic and photonic complex-valued mean-fields $\Psi_x$ and $\Psi_c$. The losses are introduced in the equations in the usual way \cite{Carusotto_2013}. In a second step, we linearize the equations of motion around the mean-field solution (Bogoliubov approximation), and thus obtain the dynamics of the small (Bogoliubov) fluctuations. These equations of motions, including the losses, can be expressed in the 
basis described earlier in a matrix form as
\begin{equation}
\frac{\partial \mathbf{\hat{A}}}{\partial t}=M_B\mathbf{\hat{A}},
\end{equation}
where $M_B$ is the Bogoliubov Matrix, and $\mathbf{\hat{A}}=[\hat {a}_1...\hat{a}_N,\hat{b}_1...\hat{b}_N,\hat a_1^\dagger...\hat{a}_N^\dagger,\hat{b}_1^\dagger...\hat{b}_N^\dagger]^T$ is a column vector of excitonic and photonic fluctuations operators where the index refers to the $N$ elements of the Bessel function basis considered. The Bogoliubov levels are obtained by diagonalizing $M_B$. 

Formally, the Bogoliubov eigenstates consists in linear superpositions of the polaritonic transverse states, however, since the energy scale of the nonlinear terms are comparable or smaller than the splitting between two consecutive states (such as $1s-1p$ or $1p-1d$), the Bogoliubov states remain dominated by a single eigenstate, and the labeling ($1s$, $1p$, $1d$) remains approximately valid. The resulting levels and labeling, obtained from fitting this model to the measured Bogoliubov spectrum is shown in Fig.\ref{fig:fig1}.(b). The fitting parameters are the two characteristic nonlinear energies $g_{s}|\Psi_x|^2\simeq 0.7$\,meV and $g_{xx}|\Psi_x|^2\simeq 0.0\,$meV. Note that the negligible value of the latter is consistent with previous observations in an unconfined 2D geometry \cite{Richard_2025}, but due to the larger uncertainties on $\omega_x$ and to the quadratic approximation for $H_V$ that are specific of the micropillar microcavity system, a quantitative analysis of the ratio of $g_{xx}$ and $g_{s}$  would require a dedicated experimental work which is beyond the scope of this work.
 
\subsection{Single-mode Bogoliubov model of correlated time-ordered photon pairs}

\subsubsection{Simpler case of vanishing bandwidth}
\label{metd:simple_th}

The Hamiltonian $\hat H_0$ of a single mode Kerr resonator, characterized by a linear frequency $\omega_0$, two-body interaction rate $g$, and a quasi-resonant drive $F$ reads:
\begin{eqnarray}
\label{eq:H_0}
    H_0&=&\hbar(\omega_0-\omega_{las}) \hat a_0^\dagger \hat a_0 \nonumber \\
    &+&(\hbar g/2) \hat a_0^\dagger\hat a_0^\dagger \hat a_0\hat a_0+\hat a_0^\dagger F+\hat a_0 F^*,
\end{eqnarray}
expressed in the frame rotating at $\omega_{las}$, the resonant drive frequency, and where $\hat a_0$ is the intracavity photonic operator. In the weakly interacting regime, we can take the Bogoliubov approximation, that consists in splitting the intracavity field into two components: a large mean-field $\psi_0$ on one hand, and the photonic fluctuations $\hat a$ on the other hand . This approximation allows deriving a steady-state equation for the mean-field $\psi_0$ \cite{Carusotto_2013}, and a quadratic Hamiltonian for the fluctuations, that reads:
\begin{eqnarray}
\label{eq:H_f}
    H_f&=&\hbar(\omega_0-\omega_{las}) \hat{a}^\dagger \hat{a} \nonumber \\
       &+&\hbar g |\psi_0|^2\hat{a}^\dagger\hat{a}+(\hbar g/2)(\psi_0^2\hat{a}^\dagger\hat{a}^\dagger+\psi_0^{*2}\hat a \hat a).
\end{eqnarray}
In the Heisenberg picture, the equation of motion for the fluctuations operator $\hat{a}$ can then be derived as
\begin{eqnarray}
    \frac{i\partial \hat{A}}{\partial t}=M_B \hat{A},
    \label{eq:EOM_a}
\end{eqnarray}
where $\hat{A}=[\hat{a},\hat{a}^\dagger]^T$, and
\begin{equation}
    M_B=\begin{pmatrix} -\delta+2g\psi_0^2-i\gamma & g\psi^2 \\  -g\psi^{*2} & \delta-2g\psi_0^2-i\gamma  \end{pmatrix},
\end{equation}
where $\hbar\delta_{las}=\hbar\omega_{las}-\hbar\omega_0$, and the loss rate $\gamma$ has been added in an effective way, that can be rigorously justified \cite{Carusotto_2013}. $M_b$ is often referred to as the Bogoliubov Matrix \cite{Carusotto_2013}. Its diagonalization results in the Bogoliubov excitation eigenstates, which is used as a preferred fluctuations representation basis in this work. The eigenstates yield the Bogoliubov transformation $\hat \beta=u\hat a-v\hat a^\dagger$, where $\hat \beta$ is a Bogoliubov fluctuation operator. The fluctuation eigenfrequency reads
\begin{equation}
    \omega_{B}=\sqrt{(-\delta\omega+2g|\psi_0|^2)^2-(g|\psi_0|)^2},
\end{equation}
where $(u,v)$ (normalized as $|u|^2-|v|^2=1$ to ensure the bosonic commutation rule) are the Bogoliubov coefficients. We assume for simplicity that $u,v$ are real, which is well verified in the regime of our experiment; generalization to complex $u,v$ comes with no subtleties. 

The photon fluctuations equation of motion can then be rewritten as a function of the Bogoliubov operator as
\begin{equation}
\hat a(t) \simeq u \hat{\beta} e^{-i\omega_B t}+ v \hat{\beta}^\dagger e^{i\omega_B t} 
\label{eq:EOM_a_of_beta_andt}
\end{equation}
where we have assumed $\gamma\ll\omega_B$. This readily Fourier transforms into
\begin{equation}
\tilde{a}(\omega) =  u \hat{\beta} \delta(\omega-\omega_B)+ v \hat{\beta}^\dagger \delta(\omega+\omega_B).
\end{equation}

Experimentally-relevant observables involve the photons emitted by the system into the extracavity continuum (of operator $\hat\alpha(t)$), that are weakly coupled to the intracavity field $\hat a(t)$ via the linear input-output relation $\hat \alpha (t) = \hat \alpha_{\rm in}(t) -i \sqrt{\gamma}\hat a(t)$, where $\hat \alpha_{\rm in}(t)$ describe the input extracavity photon vacuum fluctuations. 

Owing to the high photons energy, the system does not experience input thermal photons, so that $\hat \alpha_{\rm in}(t)$ can be simply ignored. In practice we thus have a simple linear relationship $\hat\alpha(t) \propto \hat a(t)$ between the intra- and extracavity photons. Note that this derivation is also appropriate for the case in which photons are dressed with electron-hole pairs into exciton-polaritons provided that $\gamma$ is simply corrected appropriately by the photonic Hopfield coefficients \cite{Carusotto_2013}.

For the frequency-resolved correlations that we are interested in, we consider two detectors of extracavity photons, each with a narrow frequency bandwidth acceptance $\Gamma_{bw}\ll\omega_B$, such that the first one detects only photons emitted at frequency $\omega_B$, and the second one only photons at $-\omega_B$ (where these frequencies are given as relative to the mean-field one). At these two frequencies, the previous expression yields 
\begin{eqnarray}
      \hat a(\omega_B)   &=&  u \hat{\beta} \equiv \hat{a}_N \\       
      \hat a(-\omega_B)  &= & v \hat{\beta}^\dagger \equiv \hat{a}_G,
    \label{eq:a_of_b}
\end{eqnarray}
where $\hat a(\omega_B)=\int_{\omega_B\pm\Gamma_{bw}/2} \mathrm{d}\omega'\tilde{a}(\omega')$, and $\hat{a}_N\propto\hat{\alpha}_N$ and $\hat{a}_G\propto\hat{\alpha}_G$ are defined as the \textit{Normal} and \textit{Ghost} photons operators.

Using the fact that $\hat \alpha\propto\hat a$, this analysis allows us to determine the normalized correlation of a normal-ghost emitted-photon pair. In the case where the Normal photon is emitted first reads, the correlation reads $\mathcal{G}^{(2)}_{NG}=\langle \hat{a}^\dagger_N \hat{a}^\dagger_G \hat{a}_G \hat{a}_N\rangle/(n_G n_N) \propto \langle \hat{\beta}^\dagger\hat{\beta}\hat{\beta}^\dagger\hat{\beta} \rangle$, while in the opposite order, i.e. when the Ghost photon is emitted first, it reads 
$\mathcal{G}^{(2)}_{GN}=\langle \hat{a}^\dagger_G \hat{a}^\dagger_N \hat{a}_N \hat{a}_G\rangle/(n_G n_N)\propto \langle \hat{\beta}\hat{\beta}^\dagger\hat{\beta}\hat{\beta}^\dagger \rangle$. $n_{G,N}\equiv\langle \hat a_{G,N}^\dagger \hat a_{G,N}\rangle$ is the number of Ghost/Normal photons.

These two correlations thus have a different magnitude, depending on the photons order. They can be expressed as a function of the Bogoliubov excitation number $n_b=\langle\hat{\beta}^\dagger\hat{\beta}\rangle$ as
\begin{eqnarray}
    \mathcal{G}^{(2)}_{NG} \simeq 1+\frac{n_{b}}{1+n_{b}}\\
  \mathcal{G}^{(2)}_{GN}  \simeq 1+\frac{1+n_{b}}{n_{b}},
\end{eqnarray}

where we have assumed that the photon field is in a Gaussian state and $\langle \hat{\beta}\hat{\beta}\rangle\propto\gamma/\omega_B\simeq 0$ (see SM section II). 

\subsubsection{Quantitative model accounting for a finite time and bandwidth resolution}

\label{metd:quantitative_th}

To derive a more quantitative model, which is valid for realistic values of $\gamma$, as well as the detection bandwidth $\gamma_{bw}$, and that includes explicitly the delay $\tau$ between the two photon detection events, we expand the theoretical arguments discussed above into an input-output theory. In the parameter regime of our working point, we the Bogoliubov fluctuations decay rate simply matches the cavity photon decay rate, i.e.  $\gamma_b = \gamma$.

The finite frequency resolution of the detection is accounted for by considering that each detector has a spectrally-dependent response function $f_r(\omega)$ characterized by its bandwidth $\Gamma_{bw}$ and central transmission frequency $\omega_r$. The input-output relation thus becomes $\hat{\alpha}_{r}(\tau) =  \hat{\alpha}_{\rm in}(\tau) -i\sqrt{\kappa} \int_{-\infty}^\tau dt \tilde{f}_r(\tau-t)\,\hat a(t) $, where $\tilde{f}_r(t) = e^{-i\omega_r t} e^{-\Gamma_{bw} t}$ is the response function in the time domain. The Normal-Ghost photon coincidence rate can thus be expressed as:
\begin{equation}
I^{(2)}_{NG}(\tau)=\left\langle {\cal T}_-[\hat{\alpha}_N^\dagger(0) \hat{\alpha}^\dagger_{G}(\tau)]{\cal T}_+[\hat{\alpha}_{G}(\tau)\hat{\alpha}_N(0)   ]\right\rangle,
\end{equation}
where ${\cal T}_+$ (resp. ${\cal T}_-$) orders the $a(t)$ operators with the latest time on the left (resp. right) \cite{delValle_2012}.

Assuming as before a gaussian states for the photon field, and in the limit where $\langle\hat{\beta}\hat{\beta}\rangle\simeq 0$, and $\omega_B \gg (\Gamma_{bw},\gamma_b)$, the resulting Normal-Ghost correlation function $g^{(2)}_{NG}(\tau)=I^{(2)}_{NG}(\tau)/[I_N I_{G}]$ can be derived explicitly as (see SI section III.A for the derivation)
\begin{widetext}
\begin{equation}
    g^{(2)}_{N,G}(\tau) = 1 + \frac{n_b}{n_b+1} \frac{\Gamma^2}{(\Gamma - \gamma)^2}\left[ e^{-\gamma \tau}  - \frac{e^{-\Gamma \tau}}{2\Gamma n_b}(\gamma - \Gamma + 2\gamma n_b)\right]^2
\end{equation}
\begin{equation}
    g^{(2)}_{G,N}(\tau) = 1 + \frac{n_b+1}{n_b} \frac{\Gamma^2}{(\Gamma - \gamma)^2}\left[ e^{-\gamma \tau}  - \frac{e^{-\Gamma \tau}}{2\Gamma (n_b+1)}(\Gamma - \gamma + 2\gamma(n_b+1))\right]^2
\end{equation}
\end{widetext}
These expressions are equivalent to Eqs.~(\ref{eq:g2_of_tau_single_mode}-\ref{eq:xi_of_n}) in the main text. 

\bibliography{Main_bib}

\end{document}


\title{SUPPLEMENTAL MATERIAL: Emission of time-ordered photon pairs from a coherently-driven Kerr microcavity}

\maketitle

\section{Experiment}

\subsection{Experiment details}

\begin{figure}[h!]
\includegraphics[width=0.7\textwidth]{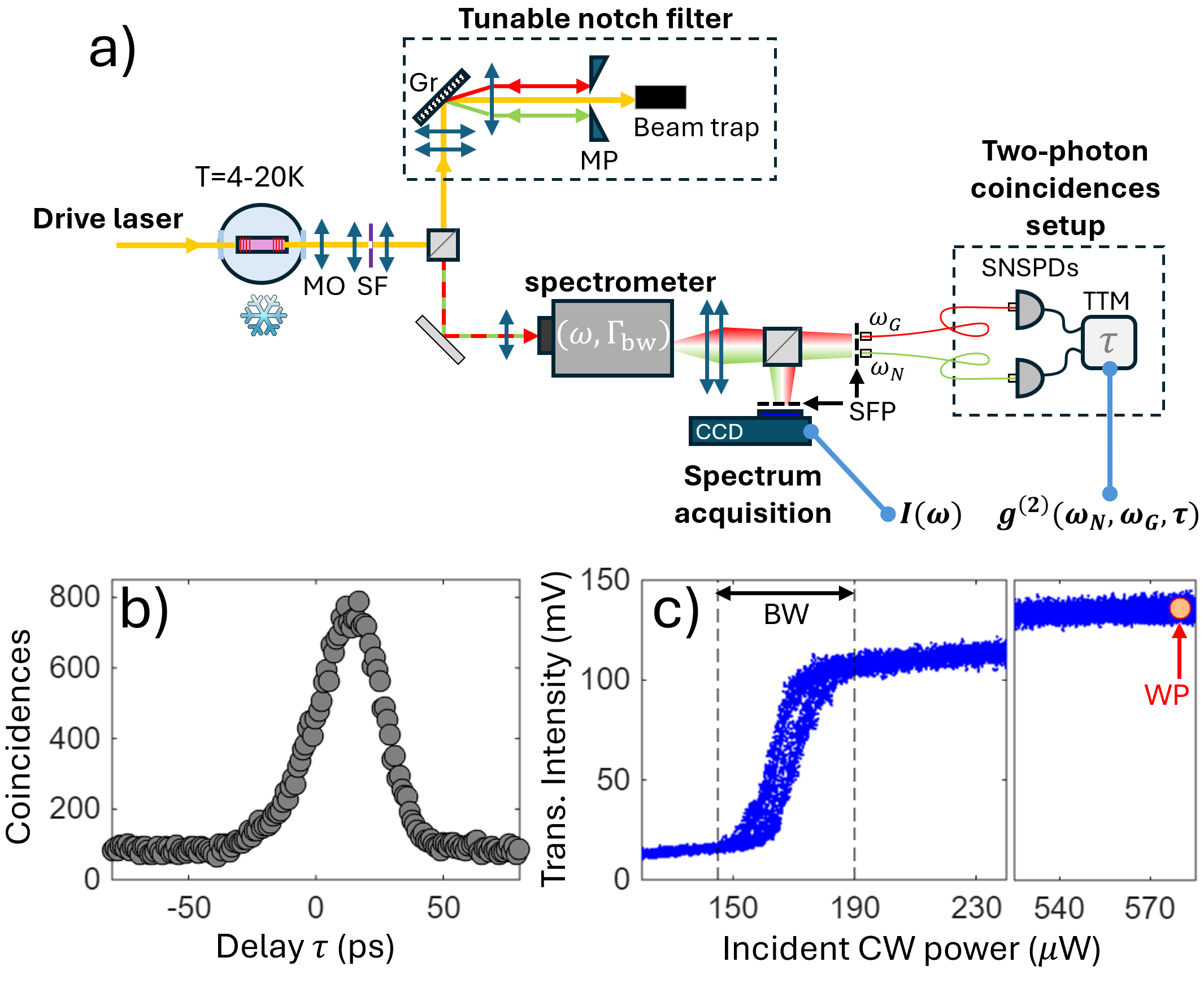}
\caption{(a) Sketch of the experimental setup. The light is collected by a $0.5$ NA microscope objective (MO), and spatially filtered (SF). The tunable custom-made notch filter involves a Grating (Gr), and split mirrors-pair assembly (MP). The filtered light is then passed through a $f=550mm$ imaging spectrometer (Horiba iHR-550). Its output spectrally-resolved focal plane (SFP) is sent either to a CCD to collect the spectral information $I(\omega)$, or into the two-photon coincidence detection setup. In the latter, the input-end of two optical fibers collect the light at the two chosen frequencies ($omega_N$ and $\omega_G$) in the SFP. Each fiber is connected to a superconducting nanowire single photon detector (SNSPD). The pulse generated at each single photon detection event is then sent into a high-resolution time-tagger module (TTM, Swabian Instrument) for delay $\tau$ measurement. (b) Raw two-photon events delay histogram corresponding to Fig.1.(d) in the main text ($7\,\mu$m micropillar, $1p-$N/$1p-$G cross correlation, $T=4.1\,$K,integration time $T_{int}=19\,$mn, Bin size: $1\,$ps, pair detection rate per bin is $0.075\,$Hz. (c) Hysteresis curve $|\psi_0|^2(I_{las})$) of the transmitted intensity $\propto|\psi_0|^2$ (as a measured voltage on a photodiode), as a function of the driving laser intensity ($I_{las}$, in $\mu$W just before impinging the micropillar) for the $7\,\mu$m micropillar ($T=4.0\,$K). 'BW' shows the bistability window. 'WP' shows the working point chosen for the NG correlation measurements. For all micropillars, the WP is chosen far on the high $I_{las}$ side of the BW for $\psi_0$ stability purpose. The BW is slightly increasing (decreasing) for larger (smaller) diameter micropillar.}
\label{fig:S1}
\end{figure}

The experimental setup that we implemented and used in this work to reject the large mean-field contribution, and measure the dim fluctuations photoemission with space, frequency and time resolution, is shown in Fig.\ref{fig:S1}.(a). The micropillar, kept in a temperature tunable cryostat ($T=[4-20]\,$K) is driven from the left by coherent light. The transmitted light is collected from the right side by a microscope objective ('MO' in Fig.\ref{fig:S1}.(a)), and spatially filtered ('SF' in Fig.\ref{fig:S1}.(a)) to reject the light emitted by the micropillar edges, that are of lower crystalline quality, and hence involves more emitters from defect states, due to the etching process.

the total emission is sent first into a custom-made tunable notch filter which is designed on the same principle as a double subtractive spectrometer in Raman spectroscopy. It rejects a $150\,\mu$eV spectral band centered on the laser frequency, and transmit the remainder of the photoemission with a flat response over $10\,$meV on both side of the laser frequency. The filter central frequency is continuously tunable in terms of center frequency, by finely turning the grating ('Gr' in Fig.\ref{fig:S1}.(a)), and in terms of rejection bandwidth by tuning the pass-through gap between the split mirror pair assembly ('MP' Fig.\ref{fig:S1}.(a). This filter is image-conserving.

The light resulting field, stripped from the mean-field, is then sent into an imaging spectrometer for spectral dispersion. We thus obtain a spatially and spectrally-resolved image in the spectrometer focal plane ('SFP' in Fig.\ref{fig:S1}.(a)). The spectrometer input slit aperture, is also used to finely adjust the spectral resolution $\Gamma_{bw}$, and hence the time-resolution in the ensuing two-photon coincidence measurements. We typically used a slit opening yielding $\hbar\Gamma_{bw}=150\,\mu$eV, i.e. low enough to separate the Normal $1p$ mode photoemission, from that of the neighboring $1s$ and $1d$ modes, and large enough to retain a high time-resolution. For the measurement of steady-state fluctuation spectroscopy $I_B(x,\omega)$ (Fig.1.b of the main text), this image is conjugated onto a sensitive CCD camera ('CCD' in Fig.\ref{fig:S1}.(a)). 
 
\subsection{2-photons coincidence detection setup}

For the measurement of two-photons coincidence with frequency, space, and time resolution,the SFP is conjugated in another plane, in which two single-mode optical fibers can each pick a different position in space and frequency, within a collection area $C_1(x_a,\omega_a)$ and $C_2(x_b,\omega_b)$ (see main text) where $\omega_{a,b}$ (e.g. $a=1p,N$ and $b=1p,G$) refers to the probed transverse mode frequencies collected by each fibers. Spatially, $(x_a,x_b)$ are chosen to coincide with an anti-node of the field mode. In agreement with the single mode theory, as long as the background photons or other modes photons have a negligible contribution at the probed frequency $\omega_{N,j}$ and $\omega_{P,j}$, $g^{(2)}_{NG,j}(\tau)$ does not depend on the positions $x_a$ and $x_b$.

The other end of each fiber is connected to a superconducting nanowire single-photon detector ('SNSPD' in Fig.\ref{fig:S1}.(a)). The resulting two channels of photo-detection events are sent into a time-tagger module ('TTM' in Fig.\ref{fig:S1}.(a)) that measures accurately the delay between each detector clicks ($1.4\,$ps jitter). Considering the $12\,$ps FWHM jitter of each SNSPD and the TTM jitter, the delay resolution is $17\,$ps FWHM, which corresponds to a 
\begin{equation}
\Delta\tau_{1\sigma}=7.3\,\mathrm{ps} 
\end{equation}
$1\sigma-$confidence interval for the measured delay ($14.5\,$ps wide $2\sigma-$confidence interval). If we include the typically chosen spectrometer bandwidth $\Gamma_{bw}\simeq 150\,\mu$eV, the $1\sigma-$confidence interval increases to $8.5\,ps$. This is to be compared with the characteristic (population) decay time $\tau_{1p}=11.5\,$ps (i.e. $\hbar\gamma=58\,\mu$eV full-width-at-half-maximum spectral bandwidth) of the $1p$ mode in the $7\mu$m diameter micropillar at $T=4.0\,$K. The raw coincidence counts histogram $N^{(2)}(\omega_{N,1p},\omega_{G,1p},\tau)\equiv N^{(2)}_{NG,1p}(\tau)$ used to determine the normalized correlation $g^{(2)}_{NG,1p}(\tau)$ shown in main text's Fig.2.c ($7\,\mu$m micropillar, $T=7\,$K), is shown in Fig.\ref{fig:S1}.(b). It was obtained in a $T_{int}=19\,$mn integration time.

From the number of two-photon coincidences $N^{(2)}_{NG,j}(\tau)$ counted in each time-bin centered on $\tau$, and of bin width $\Delta\tau=1\,$ps, and from the photon detection rate $R_{N,G}$ received by each SNSPD, one obtains the normalized two-photon cross-correlation function 
\begin{equation}
g^{(2)}_{NG,j}(\tau)=N^{(2)}_{NG,j}(\omega_N,\omega_G,\tau)/(R_NR_GT_{int}\Delta\tau),  
\end{equation}
where $T_{int}$ is the integration time, and $N^{(2)}_{NG,j}=I^{(2)}_{NG,j}(\tau)T_{int}$, where $I^{(2)}_{NG,j}(\tau)$ is the two-photon coincidence rate in each time-bin $\tau$.

\subsection{Bistability characteristics, and choice of working point}

As explained in the main text, the system is tuned into the bistable regime by choosing a laser detuning $\hbar\delta_j=\hbar\omega_{las}-\hbar\omega_{j}\geq \sqrt{3}\gamma_j$, where $\hbar\delta_j=\sqrt{3}\gamma$ is the threshold of the bistability regime in the mean-field limit \cite{Carusotto_2013}. For the $7\,\mu$m micropillar for instance, we used $\hbar\delta_{las}=90\,\mu$eV$\,=\sqrt{3}\gamma_{lp,1s}\times 1.16$. The corresponding hysteresis curve $I_{tr}(P_{in})$ is shown in Fig.\ref{fig:S1}.(c). It is obtained at a $P_{in}(t)$ scanning rate of $5\,$MHz to maximize the accuracy on the size of the bistability window \cite{Rodriguez_2017} ('BW' in Fig.\ref{fig:S1}.(c)) as we want to choose a working point ('WP' in Fig.\ref{fig:S1}.(c) away from it).

Indeed the bistable regime is necessary to achieve an optimum coupling of the laser, when the total interactions energy is comparable with the driven mode linewidth $gn \gtrsim \gamma_j$ since in the high intensity, the effective laser detuning drops to $\simeq 0$, i.e. $gn+\delta\simeq 0$, so that the laser is effectively resonant. However, we do not want the fluctuations to be perturbed by stochastic jumps between the high and low intensity states, that are known to occur within the bistable window \cite{Rodriguez_2017,Fink_2018}. We thus choose a WP far above the BW (see 'WP' in Fig.\ref{fig:S1}.(c)), i.e. in a high intracavity field intensity state which is not part of a bistable pair, for the Normal-Ghost correlations measurements.

\section{Correlation measurements}

\begin{figure}[ht!]
\includegraphics[width=0.75\textwidth]{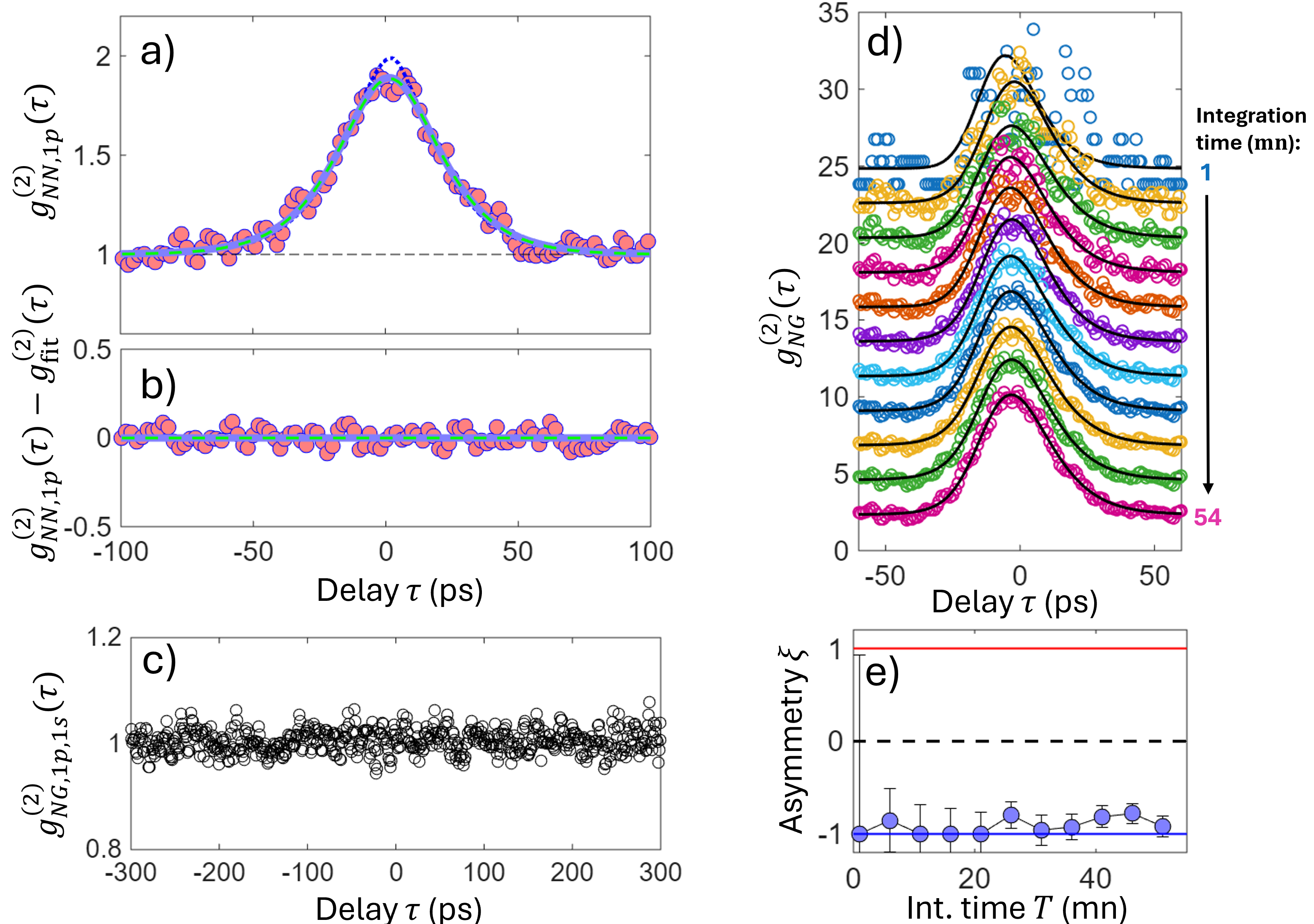}
\caption{(a) Autocorrelation measurement (red symbols) $g^{(2)}_{NN}(\tau)$ between photons picked at the same frequency $\omega_N$ of the $1p-$Normal mode ($7\,\mu$ micropillar, $T=4.0\,$K). The fit with the theory (relaxed model) is shown as a solid blue(green dashed) line. The theory and relaxed model include a Gaussian convolution to account for the experimental delay jitter of $12\,$ps (FWHM). The residual of the fit with the theory is shown in panel (b). The relaxed model fit yields an asymmetry $\xi=0.0 \pm 0.36$ ($2\sigma$ confidence interval). The deconvolved fitted relaxed model (blue dotted line) yields $max[g^{(2)}_{NN}(\tau)=1.99\pm 0.22]$ ($2\sigma$ confidence interval). (c) Measured cross-correlation $g^{(2)}_{NG,1p,1s}$ between two different modes (of the $7\,\mu$m micropillar, $T=4.0\,$K): $1p$-Normal and $1s$-Ghost (in the low frequency tail of the latter). The chosen frequencies ($\omega_{G,1s},\omega_{N,1p})$ differ from $(\omega_{G,1p},\omega_{N,1s})$ by more than both $\gamma_{1s}$ and $\gamma_{1p}$, to ensure that the measured cross-correlation involves only photons from two different modes. (d) Measured $g^{(2)}_{NG,1p}(\tau)$ as a function of the integration time $T_{int}$: $T_{int}=1\,$mn for the topmost curve, $T_{int}=54\,$mn for the bottommost curve, for the intermediate curves $T_{int}$ increases by $5.3\,$mn from top to bottom ($7\,\mu$m micropillar, $T=4.0\,$K). The asymmetry $xi$ is obtained in each curve by fitting with the relaxed model. The resulting measured $\xi(T_{int})$ is plotted in panel (e).}
\label{fig:S2}
\end{figure}

\subsection{Additional correlation measurements}

\subsubsection{Autocorrelation measurements}

According to our model, the autocorrelation of the emission of a given mode of fluctuations, i.e. $g^{(2)}_{a,j}(\tau)=g^{(2)}_{j}(\omega_j,\omega_j,\tau)$ has a peak value of $\tau$ irrespective of $\Gamma_{bw}$. It is derived explicitly in section \ref{sec_SM:theory} thereafter, Eq.(\ref{eq:g2_autocorr}). We thus measured $g^{(2)}_{NN,1p}(\tau)$ in the $1p$ mode of the $7\,\mu$m micropillar at $T=4\,$K and obtained the results shown in Fig.\ref{fig:S2}.(a). The fit with Eq.(\ref{eq:g2_autocorr}) is shown on top of the measurement in Fig.\ref{fig:S2}.(a-b) as a dashed green line. The agreement is excellent.

A fit with the relaxed model (main text Eq.(7) without enforcing Eqs.(10)) is also shown as the solid blue line to determine the max value of the correlation, and look for any asymmetry in the measurement. The fit is excellent and yields $\max[g_{NN,1p}^{(2)}(\tau)]=2\pm0.15$ upon removing the convolution by the delay measurement uncertainty (blue dotted line), as expected from the theory. The thus found asymmetry amounts to $\xi=0.14\pm0.22$, which is in full agreement with the theory $\xi=0$.

\subsubsection{Correlations between photons two different modes: $1s$-G and $1p$-N }

For the sake of completeness, we verified that the correlation between two different modes yields no correlations. Fig.\ref{fig:S2}.(c) shows $g^{(2)}_{1pN,1sG}(\tau)=g^{(2)}(\omega_{1p,N},\omega_{1s,G},\tau)$, where $\omega_{1p,N}$ is the Normal-$1p$ Bogoliubov resonance frequency, and $\omega_{1s,G}$ is picked on the red spectral tail of the $1s$ Bogoliubov resonance, where $\hbar\omega_{1s,G}=-\hbar\omega_{1p,N}+520\,\mu$eV. This data clearly shows that $g^{(2)}_{1pN,1sG}(\tau)=1$ at all delays, in this condition.

\subsubsection{Ruling-out other sources of asymmetric correlations}

In order to check that the asymmetry is not induced by the experimental setup, we performed several additional checks. First, let's keep in mind that 
\begin{itemize}
    \item 
    the correlations between asymmetry $\xi$ and sample temperature $T$ that we use to change $n_b$ is very hard to attribute to an experimental artifact.
    \item 
    The optical path used to perform the measurement of the autocorrelation $g^{(2)}_{NN,1p}(\tau)$, follows the exact same  path as for the Normal photons in the $g^{(2)}_{NG,j}(\tau)$ measurement, except at the very end when the field is split for the autocorrelation measurement, and $g^{(2)}_{NN,1p}(\tau)$ is found to be symmetric as discussed above.
    \item 
    As shown in Fig.2.(c,e) of the main text, we found that the asymmetry is swapped upon swapping the collection fibers input end. This rules out any asymmetry-inducing drifts in the detectors or TTM.  
    \item To check that the asymmetry is not appearing due to some slow uncontrolled drift occurring over integration time, we chose the conditions yielding the highest asymmetry $\xi\simeq 1$ and the highest coincidence rate at the same time, by setting the temperature to $T=4\,$K in the $6\,\mu$m, micropillar, and recorded $g^{(2)}_{NG,1p}(\tau)$ every minute during an integration time of $T_{int}=54\,$mn. 
    
    The resulting measured correlation functions are shown in Fig.\ref{fig:S2}(d) (one every $5\,$mn from top to bottom, starting at $T_{int}=1\,$mn for the topmost), and the asymmetry contained in the function is extracted by fitting it with the relaxed model. The obtained values of $\xi(T_{int})$ are shown in Fig.\ref{fig:S2}(e): as soon as the first $T_{int}=6\,$mn, a non-zero asymmetry is established with a $2-\sigma$ confidence (vertical error bars) interval. $\xi\simeq 1$ is then stable throughout the whole integration time, with only the uncertainty improving.
\end{itemize}

\subsection{Estimation of the background photons fraction at the Normal and Ghost $1p$ Bogoliubov frequencies }

\begin{figure}[h!]
\includegraphics[width=0.8\textwidth]{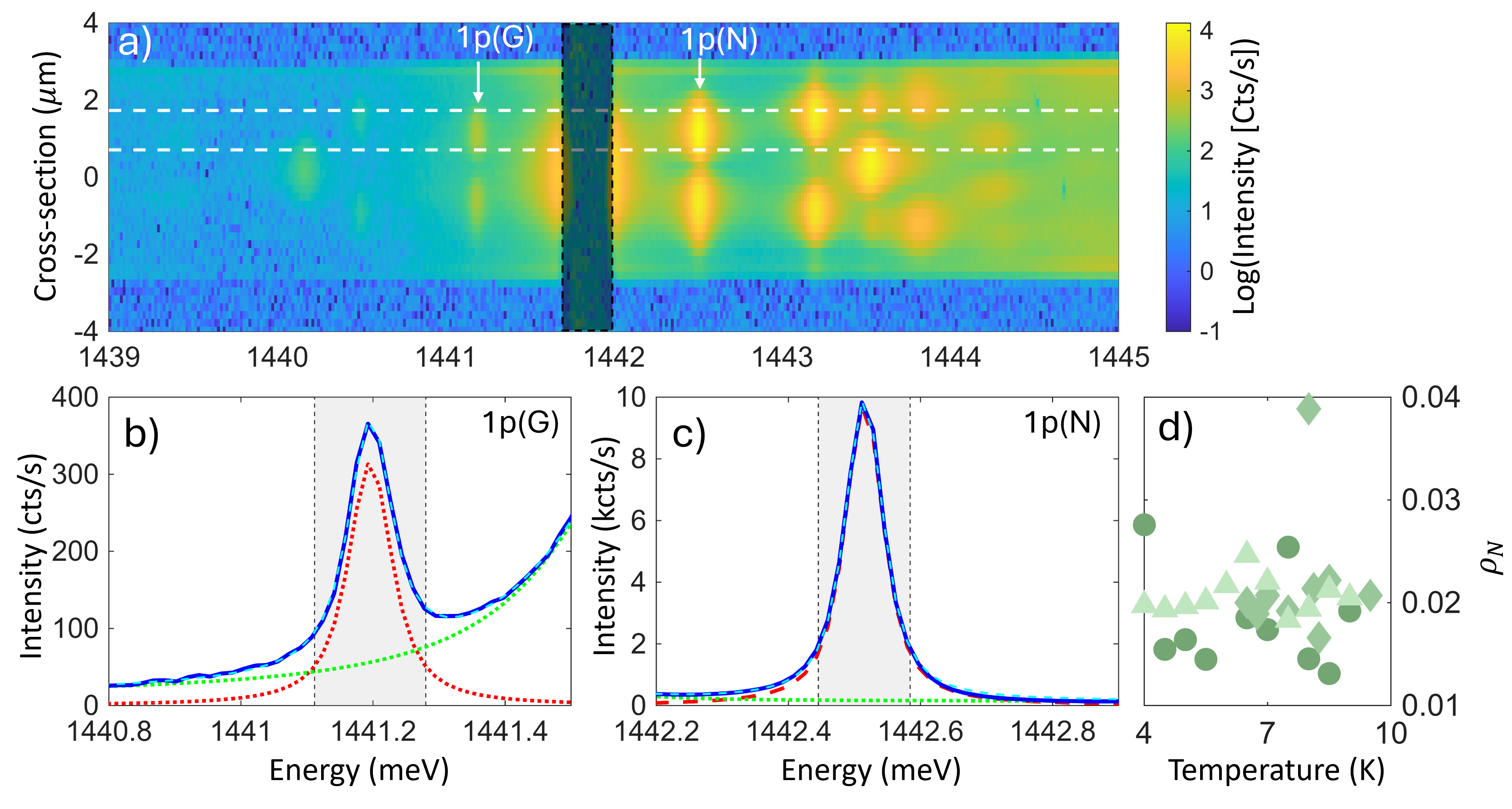}
\caption{a) Spatially-resolved Bogoliubov emission spectrum $I(\omega,x)$ in Color Log scale ($7\,\mu$m micropillar). The Normal and Ghost $1p$ Bogoliubov resonances '$1p$(N)' and '$1p$(G)', are indicated. The horizontal dashed line show the width of the spatial integration for the extracted cross-section spectrum $I_c(\omega)$. $I_{c,G}$ and $I_{c,N}$ show details of $I_c(\omega)$ focused on the Ghost (b) and Normal (c) resonances respectively (blue solid line). The fitted spectra $I_{f;N,G}$ are shown as cyan dashed line. They consist in a sum of the background contribution $I_{BG;N,G}(\omega)$ (green dotted line) and of the $1p$ Bogoliubov photons $I_{1p;N,G}(\omega)$ (red dotted line). The gray area, indicates the bandwidth in which the ratio $\rho_{G,N}$ is calculated. d) $\rho_{N}$ for the 3 experimental runs using the same conventions as in main text Fig.3.b.}
\label{fig:S3}
\end{figure}

The background contribution $\rho_{G,N}$ is determined from the measured spatially-resolved spectra $I_{\omega,x}$ such as the one shown in Fig.2.b of the main text, or the one shown in Fig.\ref{fig:S3}.(a) ($T=4\,$K) as follows. We extract a cross-section spectrum $I_c(\omega)$, spatially centered on one of the two spatial antinode of the $1p$ mode, and over a spatial width which is comparable with that collected by the fiber (cf. dashed white line in Fig.\ref{fig:S3}.(a). The thus obtained spectra are shown in Fig.\ref{fig:S3}.(b-c) in the vicinity of the Ghost ($I_{c,G}(\omega)$) and Normal ($I_{c,N}(\omega)$) resonance respectively. 

Each spectrum consists in a well-defined and well-identified peak corresponding to the $1p$ Bogoliubov resonance $I_{1p;N,G}$, on top of a slowly varying background $I_{BG;N,G}$ due to the spectral tail of the Bogoliubov $1s-$mode photons, and to a broad background probably due to bare electron-hole emission occurring around the bare exciton level. these spectra are fitted using $I_{f;N,G}(\omega)=I_{1p;N,G}(\omega)+I_{BG;N,G}(\omega)$ where a known Voigt lineshape is used for the Bogoliubov peak, and an off resonant shallow broad peak is used for the Background. The result with the different contributions is shown in Fig.\ref{fig:S3}.(b-c). The background contribution $\rho_{N,G}$ is then determined as $\rho_{N,G}=\sum_{j\in \Delta_{N,G}} I_{BG;N,G}(\omega_j)/(I_{1p;N,G}(\omega_j)+I_{BG;N,G}(\omega_j))$, and the sum runs over an interval $\Delta_{N,G}=\omega_{1p,N,G} \pm 2.5\gamma_{1p,N,G}$ centered on the Bogoliubov peak resonance $\omega_{1p,N,G}$ and extends over $2.5\times$ the resonance linewidth $\gamma_{1p;N,G}$.

\subsection{correction procedure of the measured correlation amplitude A}

A non-zero $\rho_G$ means that not only $1p$ photons but also photons from other sources are both contributing to pair counts, a situation that deviates from the single mode case. Fortunately, the fraction of non-$1p$ photons at the normal frequency amount to $\rho_N<4\%$ in the interval $T=[4,10]\,$K (see SI section I for the measurement). As a result, the Ghost background photons only contribute accidental pairs with the Normal photons and $\rho_G$ only leads to a correction of the normalization factor in $g^{(2)}_{NG,1p}(\tau)$ by a factor $(1-\rho_G)^{-1}$. We can thus apply this correction to determine the $1p$-only correlation amplitude in the measurement $A_{Mc}=A_{M}/(1-\rho_G)$, while the measured asymmetry $\xi_M$ needs no correction. 

\section{Theory}
\label{sec_SM:theory}

\subsection{Derivation of $g^{(2)}_{NG}(\tau)$ with finite bandwidth and time resolution}

The correlation function between two photons of frequencies $\omega_s$ and $\omega_{s'}$, emitted by a single mode of Bogoliubov excitations reads $g^{(2)}_{s,s'}=I^{(2)}_{s,s'}/[I_s I_{s'}]$, where $I_s=\langle \hat{\alpha}_s^\dagger(\tau) \hat{\alpha}_s(\tau)\rangle$, is the $\omega_s$-photon rate, which is a time-independent quantity in the steady-state. Assuming that the two filters have an equal bandwidth $\Gamma_s=\Gamma_{s'}=\Gamma_{bw}$, $I^{(2)}_{s,s'}$ can be expressed as
\begin{equation}    
    I^{(2)}_{s,s'}(\tau) = \left\langle {\cal T}_- [\hat{\alpha}^\dagger_s(T) \hat{\alpha}^\dagger_{s'}(T+\tau)] {\cal T}_+ [\hat{\alpha}_{s'}(T+\tau)\hat{\alpha}_s(T)] \right\rangle,
\end{equation}
where $\hat{\alpha}_{s}(\tau) =  \hat{\alpha}_{\rm in}(\tau) -i\sqrt{\kappa} \int_{-\infty}^\tau dt f_s(\tau-t)\,\hat a(t) $, $\kappa$ is the intracavity field decay rate, and $f_s(t) = e^{-i\omega_s t} e^{-\Gamma_{bw} t}$ is the filter transmission function in the time domain. $\hat{\alpha}_{\rm in}(\tau)$ describes the input quantum fluctuation field which is in the vacuum state. ${\cal T}_{\pm}$ denotes time ordering (resp. negative time ordering) for the particle operator $\hat a(t)$. In the experimental condition of steady-state, the expression does not depend on the time $T$, and hence we have:
\begin{widetext}
    \begin{equation}
        I^{(2)}_{s,s'}(\tau) \propto \int_{-\infty}^0 dt_1 f_s(-t_1) \int_{-\infty}^0 dt_2 f_s^*(-t_2) \int_{-\infty}^{\tau} dt_1' f_{s'}(\tau -t'_1) \int_{-\infty}^{\tau} dt_2' f_{s'}^*(\tau-t'_2) \left\langle {\cal T}_- [\hat{a}^\dagger(t_2) \hat{a}^\dagger(t_2')] {\cal T}_+ [\hat a(t_1')\hat a(t_1)] \right\rangle
\end{equation}
\end{widetext}

Assuming that we are only dealing with gaussian quantum states, we can use Wick's theorem, which leads to:
\begin{widetext}
\begin{eqnarray}
    \left\langle {\cal T}_- [\hat{a}^\dagger(t_2) \hat{a}^\dagger(t_2')] {\cal T}_+ [a(t_1')a(t_1)] \right\rangle &=& \langle \hat{a}^\dagger(t_2) \hat a(t_1) \rangle \langle \hat{a}^\dagger(t_2') \hat a(t_1') \rangle \nonumber\\ &+& \langle \hat{a}^\dagger(t_2) \hat a(t'_1) \rangle \langle \hat{a}^\dagger(t_2') \hat a(t_1) \rangle \nonumber\\ &+& \left\langle {\cal T}_- [\hat{a}^\dagger(t_2) \hat{a}^\dagger(t_2')] \rangle \langle {\cal T}_+ [\hat a(t_1') \hat a(t_1)] \right\rangle.
\end{eqnarray}
\end{widetext}

We see that the fundamental building blocks of the expression for $I^{(2)}_{s,s'}$ are the particle correlations $\langle \hat{a}^\dagger(t) a(t') \rangle$ and $\langle a(t) a(t') \rangle$. We evaluate them thereafter. First, we introduce the quantities:
\begin{eqnarray}
    g_{s,s'}(\tau) &=& \langle \hat{\alpha}^\dagger_{s'}(\tau) \hat{\alpha}_s(0) \rangle \\
    I_s &=& g_{s,s}(0) \\
    h_{s,s'}(\tau) &=& \langle {\cal T}_+[\hat{\alpha}_{s'}(\tau) \hat{\alpha}_s(0)] \rangle 
\end{eqnarray}
Notice that time ordering appears in $h_{s,s'}$ but not in $g_{s,s'}$. 
Then we have:
\begin{equation}
    I^{(2)}_{s,s'}(\tau) = I_s I_{s'} + |g_{s,s'}(\tau)|^2 + |h_{s,s'}(\tau)|^2
\end{equation}
and for $g^{(2)}$:
\begin{equation}
    g^{(2)}_{s,s'}(\tau) = \frac{I^{(2)}_{s,s'}(\tau)}{I_s I_{s'}} = 1 + \frac{|g_{s,s'}(\tau)|^2 + |h_{s,s'}(\tau)|^2}{I_s I_{s'}}
\end{equation}

\subsubsection*{Analytical derivation of $g^{(2)}_{s,s'}(\tau)$}

We then evaluate the two-photon correlations. We introduce the quantities:
\begin{eqnarray}
    g(\tau) &=& \langle \hat{a}^\dagger(\tau) \hat a(0) \rangle \\
    h(\tau) &=& \langle \hat a(\tau) \hat a(0)\rangle
\end{eqnarray}
Throughout the calculation, we assume that $\langle \hat \beta\hat \beta \rangle = 0$. This approximation can be well justified in the limit where $\omega_B\gg\gamma$ when adding a dissipative term coupled to the photon operator $\hat a$. We also introduce a finite linewidth $\gamma$ to the mode at frequency $\omega_B$. 

Using $\hat a(t)=[u\hat \beta e^{-i\omega_B t} + v\hat\beta^\dagger e^{i\omega_B t}]e^{-\gamma t}$, we find for $\tau \ge 0$:
\begin{eqnarray}
    g(\tau) = e^{-\gamma \tau}[u^2 \langle \hat \beta^\dagger \hat \beta \rangle e^{i\omega_B \tau} + v^2 \langle \hat \beta \hat \beta^\dagger \rangle e^{-i\omega_B \tau}] \\
    h(\tau) = uv e^{-\gamma \tau}[\langle \hat \beta^\dagger \hat \beta \rangle e^{i\omega_B \tau} + \langle \hat \beta \hat \beta^\dagger \rangle e^{-i\omega_B \tau}]
\end{eqnarray}
Notice that we have $g(-\tau) = [g(\tau)]^*$, and that we do not need $h(-\tau)$ due to the time ordering in evaluating $h_{s,s'}(\tau)$. 

We now sketch the computation for $h_{s,s'}(\tau)$:
\begin{widetext}    
\begin{eqnarray}
    h_{s,s'}(\tau) &=& \int_{-\infty}^0 dt f_s(-t)  \int_{-\infty}^\tau dt'  f_{s'}(\tau - t') \langle {\cal T}_+[\hat a(t') \hat a(t)] \rangle \\
    &=& \int_{-\infty}^0 dt f_s(-t) \left(
    \int_{t}^\tau dt'  f_{s'}(\tau - t') \langle \hat a(t') \hat a(t) \rangle +
    \int_{-\infty}^t dt'  f_{s'}(\tau - t') \langle \hat a(t) \hat a(t') \rangle 
    \right) \\
    &=& \int_{-\infty}^0 dt f_s(-t) \left(
    \int_{t}^\tau dt'  f_{s'}(\tau - t') h(t'-t) +
    \int_{-\infty}^t dt'  f_{s'}(\tau - t') h(t-t')
    \right)
\end{eqnarray}
We notice that $h(t)$ is a sum of terms with temporal dependence $e^{(\pm i\omega_B-\gamma) \tau}$, and that $f_s(t)=e^{-i\omega_s t} e^{-\Gamma t}$. Evaluating $h_{s,s'}(\tau)$ is then a straightforward exercise of integrating exponential functions. In particular, a term $e^{(i\omega_B-\gamma) \tau}$ in $h(\tau)$ gives for $h_{s,s'}(\tau)$ a contribution:
\begin{eqnarray}
    h_{s,s'}^{\rm (term~+\omega_B)}(\tau) &=& \frac{e^{-\Gamma\tau} e^{-i\omega_{s'}\tau}}{2\Gamma + i\omega_s + i\omega_{s'}}\left[\frac{1}{\Gamma + \gamma + i\omega_{s'} - i\omega_B} - \frac{1}{\Gamma - \gamma + i\omega_{s'} + i\omega_B} \right] \nonumber\\ 
    && + \frac{e^{-\gamma \tau} e^{i\omega_B \tau}}{(\Gamma - \gamma + i\omega_{s'} + i\omega_B)(\Gamma + \gamma + i\omega_s - i\omega_B)}
\end{eqnarray}
Therefore, gathering both $e^{\pm i\omega_B}$ contributions, the final result for $h_{s,s'}(\tau)$ reads:
\begin{eqnarray}
    h_{s,s'}(\tau) &=& e^{-\Gamma \tau} e^{-i\omega_{s'}\tau} \frac{uv}{2 \Gamma + i\omega_s + i\omega_{s'}} \left[ 
         \langle \hat \beta^\dagger \hat \beta \rangle\left(\frac{1}{\Gamma + \gamma + i\omega_{s'} - i\omega_B} - \frac{1}{\Gamma - \gamma + i\omega_B + i\omega_{s'}} \right) \right. \nonumber \\
        && + \left. \langle \hat \beta \hat \beta^\dagger \rangle \left(\frac{1}{\Gamma + \gamma + i\omega_B + i\omega_{s'}} - \frac{1}{\Gamma - \gamma - i\omega_B + i\omega_{s'}} \right) \right] \nonumber\\
        && + e^{-\gamma \tau} e^{i\omega_B \tau} \frac{uv \langle \hat \beta^\dagger \hat \beta \rangle}{(\Gamma - \gamma + i\omega_B + i\omega_{s'})(\Gamma + \gamma - i\omega_B + i\omega_s )} \nonumber \\
        && + e^{-\gamma \tau} e^{-i\omega_B \tau} \frac{uv \langle \hat \beta \hat \beta^\dagger \rangle}{(\Gamma - \gamma - i\omega_B + i\omega_{s'})(\Gamma + \gamma + i\omega_B + i\omega_s )}
\end{eqnarray}

Similarly, for $g_{s,s'}(\tau) =  \int_{-\infty}^0 dt f_s(-t)  \int_{-\infty}^\tau dt'  f^*_{s'}(\tau - t') \langle \hat{a}^\dagger(t') a(t) \rangle$, the final result reads:
\begin{eqnarray}
    g_{s,s'}(\tau) &=& e^{-\Gamma \tau} e^{i\omega_{s'}\tau} \frac{1}{2 \Gamma + i\omega_s - i\omega_{s'}} \left[ 
        u^2 \langle \hat \beta^\dagger \hat \beta \rangle\left(\frac{1}{\Gamma + \gamma + i\omega_B - i\omega_{s'}} - \frac{1}{\Gamma - \gamma + i\omega_B - i\omega_{s'}} \right) \right. \nonumber \\
        && + \left. v^2 \langle \hat \beta \hat \beta^\dagger \rangle \left(\frac{1}{\Gamma + \gamma - i\omega_B - i\omega_{s'}} - \frac{1}{\Gamma - \gamma - i\omega_B - i\omega_{s'}} \right) \right] \nonumber\\
        && + e^{-\gamma \tau} e^{i\omega_B \tau} \frac{u^2 \langle \hat \beta^\dagger \hat \beta \rangle}{(\Gamma - \gamma + i\omega_B - i\omega_{s'})(\Gamma + \gamma - i\omega_B + i\omega_s )} \nonumber \\
        && + e^{-\gamma \tau} e^{-i\omega_B \tau} \frac{v^2 \langle \hat \beta \hat \beta^\dagger \rangle}{(\Gamma - \gamma - i\omega_B - i\omega_{s'})(\Gamma + \gamma + i\omega_B + i\omega_s )}
\end{eqnarray}

We then evaluate these expressions at resonance for N/G or G/N correlations, namely when $\omega_s=-\omega_{s'} = \pm\omega_B$. We establish approximate expressions by assuming that $\omega_B \gg \gamma,\Gamma$, and keeping only the dominant term. We find:
\begin{equation}
    h_{N,G}(\tau) = uv e^{i\omega_{0}\tau}\left[ e^{-\Gamma \tau} \frac{1}{2 \Gamma} \left(
         - \frac{\langle \hat \beta^\dagger \hat \beta \rangle}{\Gamma - \gamma } +\frac{\langle \hat \beta \hat \beta^\dagger \rangle}{\Gamma + \gamma}  \right) + e^{-\gamma \tau} \frac{\langle \hat \beta^\dagger \hat \beta \rangle}{(\Gamma - \gamma)(\Gamma + \gamma)} \right]
\end{equation}
\begin{equation}
    h_{G,N}(\tau) = uv e^{-i\omega_{0}\tau}\left[ e^{-\Gamma \tau} \frac{1}{2 \Gamma} \left(
         \frac{\langle \hat \beta^\dagger \hat \beta \rangle}{\Gamma + \gamma } -\frac{\langle \hat \beta \hat \beta^\dagger \rangle}{\Gamma - \gamma}  \right) + e^{-\gamma \tau} \frac{\langle \hat \beta \hat \beta^\dagger \rangle}{(\Gamma - \gamma)(\Gamma + \gamma)} \right]
\end{equation}

We further notice that $g_{N,G}(\tau), g_{G,N}(\tau) \sim {\cal O}(1/(\gamma\omega_B))$, hence $|g_{s,s'}|^2 \ll |h_{s,s'}|^2$ for N/G and G/N correlations evaluated at the peak frequencies. Hence, in this limit we have the approximate expression:
\begin{equation}
    g^{(2)}_{s,s'}(\tau) = 1 + \frac{|h_{s,s'}(\tau)|^2}{I_s I_{s'}}
\end{equation}
Evaluating $I_s=g_{s,s}(0)$, we find in the limit $\omega_B \gg \gamma,\Gamma$:
\begin{eqnarray}
    I_N = g_{N,N}(0) = \frac{u^2 \langle \hat \beta^\dagger \hat \beta \rangle}{\Gamma(\Gamma + \gamma)} \\
    I_G = g_{G,G}(0) = \frac{v^2 \langle \hat \beta \hat \beta^\dagger \rangle}{\Gamma(\Gamma + \gamma)}
\end{eqnarray}

We obtain the final expression for $g^{(2)}$, valid in the limit $\omega_B \gg \gamma,\Gamma$, and for the filter frequencies $\omega_s$, $\omega_{s'}$ centered on the peak frequency $\pm \omega_B$:
\begin{equation}
    g^{(2)}_{N,G}(\tau) = 1 + \frac{\langle \hat \beta^\dagger \hat \beta \rangle}{\langle \hat \beta \hat \beta^\dagger \rangle} \frac{\Gamma^2}{(\Gamma - \gamma)^2}\left[ e^{-\gamma \tau}  - \frac{e^{-\Gamma \tau}}{2\Gamma \langle \hat \beta^\dagger \hat \beta \rangle}(\gamma - \Gamma + 2\gamma\langle \hat \beta^\dagger \hat \beta \rangle)\right]^2
\end{equation}
\begin{equation}
    g^{(2)}_{G,N}(\tau) = 1 + \frac{\langle \hat \beta \hat \beta^\dagger \rangle}{\langle \hat \beta^\dagger \hat \beta \rangle} \frac{\Gamma^2}{(\Gamma - \gamma)^2}\left[ e^{-\gamma \tau}  - \frac{e^{-\Gamma \tau}}{2\Gamma \langle \hat \beta \hat \beta^\dagger \rangle}(\Gamma - \gamma + 2\gamma\langle \hat \beta \hat \beta^\dagger\rangle)\right]^2
\end{equation}
\end{widetext}
These expressions are equivalent to Eqs.~(7-9) in the main text. 

\subsection{Derivation of the autocorrelation function $g^{(2)}_{s,s}(\tau)$}

We can also determine the autocorrelation as $g^{(2)}_{N,N}(\tau)=g^{(2)}_{G,G}(\tau)\simeq1+|g_{N,N}(\tau)|^2/I_N^2$ which yields
\begin{equation}
    g^{(2)}_{N,N}(\tau) = 1 + \frac{\Gamma^2}{(\Gamma - \gamma)^2}\left[ e^{-\gamma \tau}  - \frac{\gamma}{\Gamma}e^{-\Gamma \tau}\right]^2.
    \label{eq:g2_autocorr}
\end{equation}
remarkably, the result is independent of $n_b=\langle \beta^\dagger\beta\rangle$, and furthermore we find that $g^{(2)}_{N,N}(0)=2$ irrespective of $\Gamma$. We also note that $g^{(2)}_{N,N}(\tau)=g^{(2)}_{N,G}(\tau)=g^{(2)}_{G,N}(\tau)$ upon taking $n_b\rightarrow\infty$.

\subsection{Comparison of the fluctuations equations of motions, in the single-mode Bogoliubov regime and in the spontaneous parametric emission configuration}

\subsubsection*{single-mode Bogoliubov regime}

The minimum system Hamiltonian required to generate a single mode of Bogoliubov fluctuations involves two modes: one which is used to inject the coherent drive (labeled $p$), and the other one (labeled $0$) hosting the fluctuations. Its linear part thus reads :  
\begin{eqnarray}
\label{eq:H_0}
    H_0&=& \sum_{q=p,0} \hbar\omega_q \hat A_q^\dagger \hat A_q,
\end{eqnarray}
where $\hat A_q$ is the photon operator of mode $q$. Adding the nonlinear Kerr term, and the coherent drive $\psi_0$, which is set to be resonant with mode $p$ (accounting for the nonlinear shift) and by taking the Bogoliubov expansion around $\psi_0$, we obtain a linearized Hamiltonian describing the fluctuations in modes $p$ and $0$.  As explained in the main text, the fluctuations in these two modes can be separated from each other when the condition $|\omega_{0}-\omega_p| \gg \gamma$ is met. In this case, mode $0$ can be isolated from $p$, and the fluctuations that it is hosting are described by the Hamiltonian given in the main text's Methods section Eq.(13) and reproduced thereafter:
\begin{eqnarray}
\label{eq:H_f}
    H_B&=&\hbar(\omega_0-\omega_{las}) \hat{a}_0^\dagger \hat{a}_0 \nonumber \\
       &+&\hbar g |\psi_0|^2\hat{a}_0^\dagger\hat{a}_0+(\hbar g/2)(\psi_0^2\hat{a}_0^\dagger\hat{a}_0^\dagger+\psi_0^{*2}\hat a_0 \hat a_0),
\end{eqnarray}
where $\hat a_0$ is the $0$-mode photon fluctuations operator. The corresponding equation of motions are given in main text's Methods section Eqs.(14-15), and reproduced thereafter:

\begin{eqnarray}
    \frac{i\partial \hat{V}}{\partial t}=M_B \hat{V},
\end{eqnarray}
where $\hat{V}=[\hat{a_0},\hat{a_0}^\dagger]^T$, and
\begin{equation}
    M_B=\begin{pmatrix} -\delta_0+2g\psi_0^2-i\gamma & g\psi^2 \\  -g\psi^{*2} & \delta_0-2g\psi_0^2-i\gamma  \end{pmatrix},
\end{equation}
where $\delta_0=\omega_{las}-\omega_0$, $\gamma$ is the loss rate.

$M_B$ is diagonalizable via the Bogoliubov transformation, leading to the Bogoliubov eigenstates, characterized by the real-valued eigenvector $(u,v)$ (of positive norm \cite{Castin_2007}), and the eigenfrequency $\omega_B$. This Bogoliubov eigenmode (operator $\hat \beta$) possesses two photonic fluctuations frequency components as derived in the main tex's Methods section Eq.(18-20): 

\begin{equation}
\tilde{a}(\omega) =  u \hat{\beta} \delta(\omega-\omega_B)+ v \hat{\beta}^\dagger \delta(\omega+\omega_B)\equiv\hat a_N \delta(\omega-\omega_B)+ \hat a_G \delta(\omega+\omega_B),
\end{equation}
that corresponds to the Normal (operator $\hat a_N$) and Ghost ($\hat a_G$) levels.

\subsubsection*{Spontaneous parametric emission configuration}

The minimum system Hamiltonian required for spontaneous parametric emission (of photon pairs) involves three modes, i.e. one more: $p$, and $s$ and $i$. Excluding the drive and nonlinearity, it thus reads  
\begin{eqnarray}
    H_0&=& \sum_{q=p,s,i} \hbar\omega_q \hat A_q^\dagger \hat A_q.
\end{eqnarray}
Like in the previous case, the mode $p$ is used for the mean-field $\psi_0$ injection, while mode $s$ (signal) and $i$ (idler) are chosen on both sides of level $p$, such that, once they transform into fluctuation levels, the parametric resonance condition can be achieved, as described further.

Using again the Bogoliubov approximation, like in the previous case, we can write an effective Hamiltonian for the photonic fluctuations. In the conditions $|\omega_{s,i}-\omega_p| \gg \gamma$, $s$ and $i$ modes are well isolated from the driven mode $p$. An effective Hamiltonian can then be written for the $s$ and $i$ modes only, that reads in the frame rotating at the drive frequency:
\begin{eqnarray}
    H_f&=& \sum_{q=i,s} \hbar\delta_q \hat a_q^\dagger \hat a_q+2\hbar g|\psi|^2  (\hat a_s^\dagger \hat a_s+\hat a_i^\dagger \hat a_i) \\
    && + \hbar g\psi^2(\hat a_s^\dagger \hat a_i^\dagger+\hat a_i^\dagger \hat a_s^\dagger) \\
    && + \hbar g\psi^{*2}(\hat a_s \hat a_i+\hat a_i \hat a_s),
\end{eqnarray}
where $\delta_{i,s}\equiv\omega_{las}-\omega_{i,s}$, and $\hat a_{s,i}$ are the photonic fluctuations operators in the signal and idler modes. In this Hamiltonian, only the nonlinear terms matching the parametric resonant condition ($2\delta_p\simeq0=\delta_s+\delta_i$) have been kept. The resulting equations of motions can be split into two independent sets of equations:
\begin{eqnarray}
    \frac{i\partial \hat{V_1}}{\partial t}=M_{1} \hat{V_{1}}, \text{ and } \frac{i\partial \hat{V_2}}{\partial t}=M_2 \hat{V_{2}},
\end{eqnarray}
where $\hat{V_1}=[\hat{a_s},\hat{a_i}^\dagger]^T$, and $\hat{V_2}=[\hat{a_s}^\dagger,\hat{a_i}]^T$, and
\begin{equation}
    M_1=\begin{pmatrix} \delta_s+2g|\psi_0|^2 -i\gamma & g\psi_0^2 \\  -g\psi_0^{*2} & -\delta_i-2g|\psi_0|^2-i\gamma  \end{pmatrix} \text{and } M_2=\begin{pmatrix} -\delta_s-2g|\psi_0|^2 -i\gamma & -g\psi_0^{*2} \\  
    g\psi_0^{2} & \delta_i+2g|\psi_0|^2-i\gamma  \end{pmatrix}, 
\end{equation}
where the damping rate $\gamma$ has been introduced effectively as usual. These equations couple the 'particle' of the signal mode with the 'hole' of the idler mode, and vice-versa. This is a key difference with the single mode Bogoliubov case, in which the particle and hole component involved in the fluctuations belong to the same mode, resulting in the fact that $\hat a_N$ and $\hat a_G$ do not commute ($[\hat a_N,\hat a_G]=uv$), and hence to the fact that the $N$ and $G$ photon pairs are time-ordered. On the contrary, the signal and idler modes operators do commute by construction, so that $\langle\hat a_G^\dagger\hat a_N^\dagger\hat a_N\hat a_G\rangle=\langle\hat a_N^\dagger\hat a_G^\dagger\hat a_G\hat a_N\rangle$, i.e. at zero-time delay, the signal-idler photon pairs correlations are not time-ordered. Owing to the symmetry upon swapping $\hat a_s$ and $\hat a_i$ in $H_f$, when the parametric resonance condition is enforced, this lack of time-order in the signal and idler photon pairs is expected to hold at finite delay as well.

The nature of the fluctuations is also fundamentally different. Accounting for the nonlinear shifts, the resonant parametric condition can be expressed exactly as $\delta_s+2g|\psi_0|^2=-\delta_i-2g|\psi_0|^2$, leading to $M_2=-M_1^*$. Unlike $M_B$ earlier, neither $M_1$ nor $M_2$ are diagonizable. Attempting to do so, leads to complex eigenfrequencies and eigenvectors, and attempting to normalize the thus obtained eigenstates by enforcing a bosonic commutator $[\hat f,\hat f^\dagger]=1$ as required, leads to a vanishing norm. The nature of the signal-idler fluctuations is therefore not of the Bogoliubov type.

\bibliography{Main_bib}